\documentclass[pdflatex,sn-mathphys-num]{sn-jnl}

\usepackage{graphicx}%
\usepackage{multirow}%
\usepackage{amsmath,amssymb,amsfonts}%
\usepackage{amsthm}%
\usepackage{mathrsfs}%
\usepackage[title]{appendix}%
\usepackage{xcolor}%
\usepackage{textcomp}%
\usepackage{manyfoot}%
\usepackage{booktabs}%
\usepackage{algorithm}%
\usepackage{algorithmicx}%
\usepackage{algpseudocode}%
\usepackage{listings}%
\usepackage{subcaption}%

\theoremstyle{thmstyleone}%
\theoremstyle{thmstyletwo}%

\theoremstyle{thmstylethree}%

\begin{document}

\title[Article Title]{Advances in the Vacuum Magnetic Birefringence measurement using a pulsed Foil Coil}


\author*[1]{\fnm{Enzo} \sur{Rodriguez Alvarez}}\email{enzo.rodriguez-alvarez@lncmi.cnrs.fr}
\equalcont{These authors contributed equally to this work.}

\author[1,2,3]{\fnm{Jonathan} \sur{Agil}}\email{agiljona@uni-mainz.de,\\ present address: Exail SAS Gradignan, 27 avenue de la Poterie, 33170, Gradignan, France. jonathan.agil@exail.com}
\equalcont{These authors contributed equally to this work.}

\author[1]{\fnm{Julien} \sur{Billette}}\email{julien.billette@lncmi.cnrs.fr}
\equalcont{These authors contributed equally to this work.}

\author[1]{\fnm{Jérôme} \sur{Béard}}\email{jerome.beard@lncmi.cnrs.fr}
\equalcont{These authors contributed equally to this work.}

\author[1]{\fnm{Rémy} \sur{Battesti}}\email{remy.battesti@lncmi.cnrs.fr}
\equalcont{These authors contributed equally to this work.}

\author[1]{\fnm{Carlo} \sur{Rizzo}}\email{carlo.rizzo@lncmi.cnrs.fr}
\equalcont{These authors contributed equally to this work.}

\affil[1]{\orgdiv{LNCMI}, \orgname{UPR CNRS 3228, Univ. Toulouse, INSA, Univ. Grenoble-Alpes, EMFL}, \orgaddress{\street{143 Avenue de Rangueil}, \city{Toulouse}, \postcode{31400}, \country{France}}}

\affil[2]{\orgname{Johannes Gutenberg-Universität Mainz}, \orgaddress{\street{}, \city{Mainz}, \postcode{55128}, \country{Germany}}}

\affil[3]{\orgdiv{Helmholtz-Institut Mainz}, \orgname{GSI Helmholtzzentrum für Schwerionenforschung}, \orgaddress{\city{Mainz}, \postcode{55128}, \country{Germany}}}


\abstract{In this paper, in the framework of the {\it Bir\'efringence Magn\'etique du Vide} (BMV) experiment, we present our current apparatus in detail and we discuss its critical points. We also report the results obtained using specially designed pulsed coils called Foil Coils. Our novel measurement of the vacuum linear magnetic birefringence constant is $K_{VMB} < 1.7 \cdot 10^{-21}~\text{T}^{-2}$ at a confidence level of 68$\%$, which is an improvement with respect to our previous 2014 measurement.}




\maketitle

\section{Introduction}\label{sec1}

The questions about the nature of a vacuum, its existence, and if and how a light can propagate through it, can be traced back to the beginnings of the natural philosophy \cite{Agil2023}. The current point of view is that a vacuum is a quantum state to which magnetic and electric properties can be ascribed like a standard non-linear optical medium \cite{Battesti2013}.

The first model of a vacuum in the framework of the quantum theory has been worked out by Dirac in his seminal paper about the positron \cite{Dirac1934}. Immediately, Heisenberg understood that in a Dirac's vacuum non-linear electromagnetic interactions could be possible and in particular photon-photon scattering. From the discussions about such a novel description of a vacuum between Heisenberg and Euler, that have been reported by Heisenberg himself in his autobiography \cite{Heisenberg1971}, two very important contributions to fundamental interaction physics resulted \cite{Euler1935,Heisenberg1936}. In these works, it has been established what is called the Heisenberg-Euler Lagrangian (H-E Lagrangian in the following) which is the mathematical description of the behaviour of electromagnetic fields in a vacuum. In 1951 Schwinger obtained exactly the same Lagrangian using the new developed quantum field theory \cite{Schwinger1951} and the H-E Lagrangian  is therefore considered a milestone in the field of Quantum ElectroDynamics (QED).

In the seventies of the last century, thanks to the work of Bialynicka-Birula and Bialynicki-Birula \cite{BialinskiBirula1970}, it became clear that interactions between polarised plane waves and external fields would be possible following the H-E Lagrangian, as well. Since that, from the experimental point of view, for technical reasons, much attention has been devoted to the effect of an external magnetic field on the propagation of light in a vacuum \cite{Battesti2013}. The H-E Lagrangian predictions are that light linearly polarised is slowed down by the presence of an external field perpendicular to the light wave vector of a quantity depending on the light polarisation direction.
 
If one calls $n_{\parallel}$ the index of refraction for light polarised parallel to the magnetic field $B$ and $n_{\perp}$ the one for light polarised perpendicular to  $B$, with $\Delta n_{VMB} = n_{\parallel} - n_{\perp}$, taking also into account a higher order correction to the H-E Lagrangian calculated by Ritus in 1975 \cite{Ritus1975}, the prediction is that \cite{Battesti2013}
\begin{equation}\label{VCM}
\Delta n_{VMB} = K_{VMB}B^2 = {\frac{2\alpha^2 \hbar^3}{ 15 m_{e}^4 c^5}\left(1+ \frac{25\alpha}{4\pi}\right){\frac{B^2}{\mu_{0}}}} \sim 4.03\cdot 10^{-24} \left(\frac{B}{1\, \text{T}}\right)^2,
\end{equation}
where $\alpha$, $\hbar$, $m_{e}$, $c$ and $\mu_{0}$ are the usual fundamental constants in the ref.~\cite{Mohr2025}. 

This kind of phenomenon induced by a magnetic field is called in optics a linear magnetic birefringence. It has been first observed at the beginning of the 20$^{\text{th}}$ century. It exists in any material medium and it is also called Cotton-Mouton effect, in particular in dilute matter \cite{Rizzo1997}. Despite over a century of attempts, $\Delta n_{VMB}$ has not yet been measured \cite{Agil2023}.

From a more general point of view, the H-E Lagrangian is just one of the possible Lagrangians describing field interactions in the framework of Non-Linear ElectroDynamics where $\Delta n_{VMB}$ at the lowest order can be written as a linear combination of two parameters \cite{Fouche2016,fouche2017erratum}. A precise measurement of $\Delta n_{VMB}$ could therefore exclude alternative theories like the Born-Infeld one \cite{Born1934}, which predicts a zero linear magnetic birefringence even if fields can interact with each other  \cite{Fouche2016,fouche2017erratum}, and some other extensions of the standard model. 

On the other hand, QED predictions have been tested very precisely in the fields of bound states or isolated particles (see e.g.~\cite{Sarri2025}) and it may be argued that its predictions concerning the photon sector are correct as well. But, physics beyond the standard model can also be tested looking for Vacuum Magnetic Birefringence (VMB) since apparent optical magnetic activity or linear magnetic birefringence can be induced by the existence of particles coupling with photons in an external magnetic field \cite{Maiani1986} like the AxionLike Particles (ALP) \cite{Graham2015}. The search for these elusive particles has been the main motivation for the first experimental attempt to study the light propagation in a vacuum in the presence of a transverse magnetic field \cite{Cameron1993} following the Bialynicka-Birula and Bialynicki-Birula paper \cite{BialinskiBirula1970}.

\begin{figure}[ht!]
\centering
\includegraphics[width=1\textwidth]{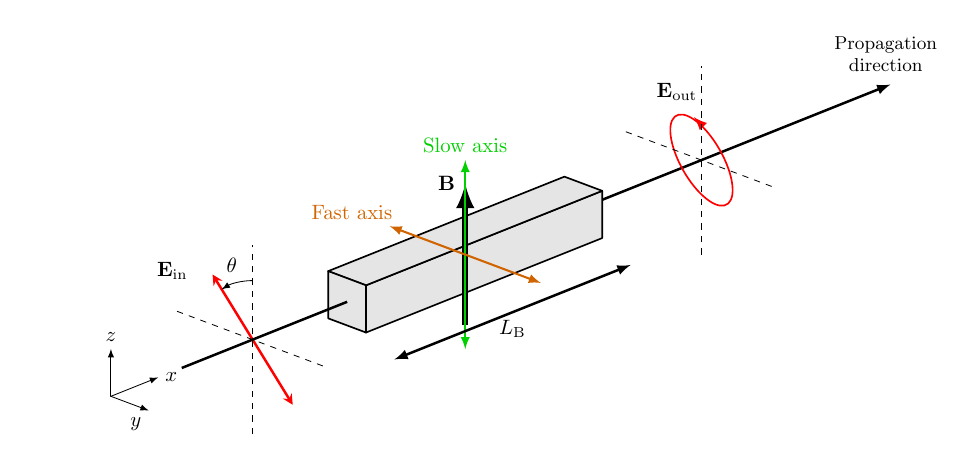}
\caption{Illustration of the principle of a linear magnetic birefringence, here induced by a transverse magnetic field $\mathbf{B}$ over a length $L_B$. The components along the two principal axes of the medium are phase-shifted; at the output, the polarisation becomes elliptical. In the case shown in our figure, the linear component of the polarisation parallel to $\mathbf{B}$ 
propagates at a slower speed with optical index $n_\parallel$ (slow axis) than the linear component of the polarisation perpendicular to it with optical index difference $n_\parallel~-~n_\perp > 0$, which transforms the incident linear polarisation into elliptical polarisation.}\label{fig1}
\end{figure}

Following the 1979 seminal paper of Iacopini and Zavattini \cite{Iacopini1979}, the $K_{VMB}$ constant is obtained by a measurement of the ellipticity $\Psi$ acquired by the initially linearly polarised laser beam in the magnetic field region whose effective length is $L_B$ (see Fig. \ref{fig1}). $\Psi$ is given by the following eq.~(\ref{PsiWOcav})
\begin{equation}\label{PsiWOcav}
\Psi = \pi \frac{L_B}{\lambda} \Delta n_{VMB},
\end{equation}
and the effect to be measured is modulated to increase the signal to noise ratio.

Up to now, the best limit reported for the vacuum magnetic birefringence constant $K_{VMB}$ has been obtained by the PVLAS group after almost 25 years of  experimental efforts \cite{Ejlli2020}: $K_{VMB} < 2.8\cdot 10^{-23}$~T$^{-2}$ at a confidence level of 68$\%$. The measurement has been performed thanks to a 2.5~T permanent magnetic field. The modulation of the effect is obtained by rotating the whole permanent magnet around its axis \cite{Ejlli2020}. 

In the framework of the BMV experiment \cite{Battesti2008}, we present in this paper our present apparatus in detail and we discuss its critical points. We also report the results obtained using specially designed pulsed coils called Foil Coils, whose commissioning has been reported in the ref.~\cite{Beard2021}. Actually, since the beginning \cite{Battesti2008}, our group has chosen to use pulsed coils which, in principle, can deliver very intense field exceeding 10~T in a few milliseconds thus providing a modulation frequency which should allow a better signal to noise ratio \cite{Berceau2012}. Our experiment is therefore hosted by the  {\it Laboratoire National des Champs Magnétiques Intenses} (LNCMI) in Toulouse, France. The data reported here have been taken between 2021 and 2025 (namely, in the following, run 2021 and run 2025, respectively) using two versions of our set-up based on the same experimental methods.

As a matter of fact, the basic principles on which our experiment is based, which have been summarized in this section, are quite challenging to be realized in a laboratory at a level allowing the measurement of $K_{VMB}$. In practice, to measure the VMB effect, one has to provide a magnetic field as high as possible, which means to deliver tens of MW of peak power, while keeping the sensitivity of an optical apparatus at levels comparable with the sensitivity of a Gravitational Wave interferometer (see e.g.~\cite{Hartman2017} and references within).

Our novel measurement of the vacuum linear magnetic birefringence constant is $K_{VMB} < 1.7 \cdot 10^{-21}\ \text{T}^{-2}$ at a confidence level of 68$\%$. This result of ours is an improvement with respect to our previous one \cite{Cadene2014} obtained by using a different coil, called Xcoil \cite{Cadene2014}.

In section \ref{sec2}, we present our experimental method ; in section \ref{sec3} we review the main facts concerning our runs 2021 and 2025 and report our results obtained using Foil Coils ; in section \ref{sec4}, finally, we state our conclusions and discuss perspectives.

\section{Experimental method}\label{sec2}

\begin{figure}[ht!]
\centering
\includegraphics[width=1\textwidth]{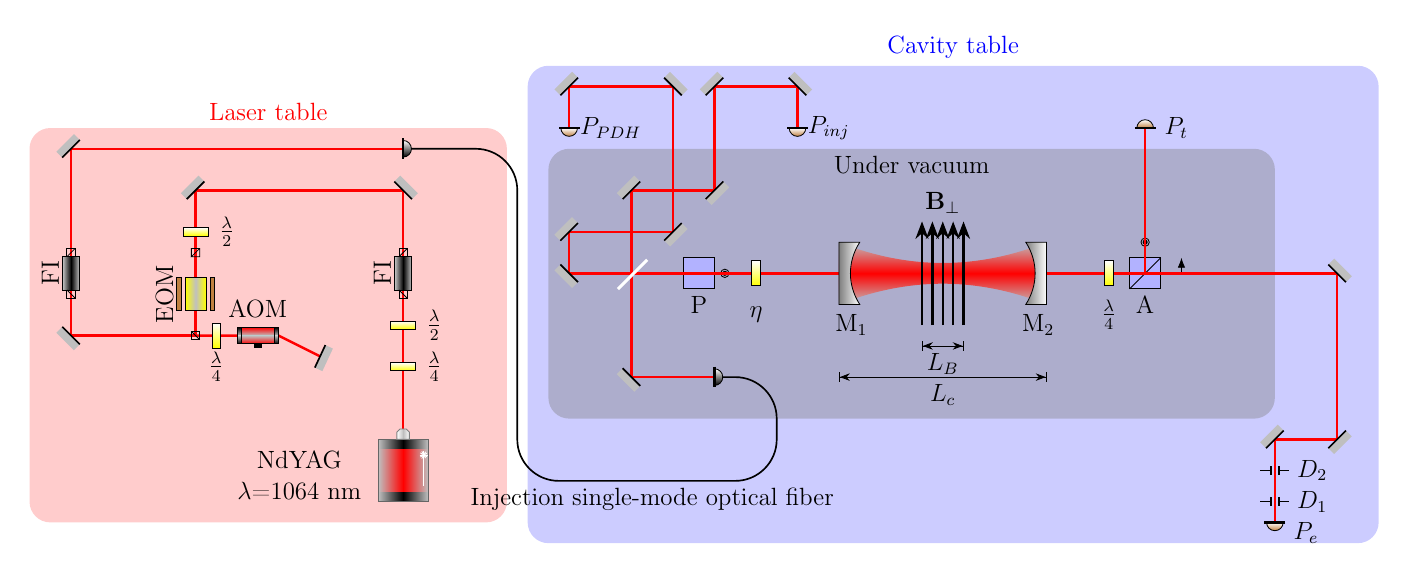}
\caption{Simplified optical scheme of the BMV experiment. On the laser table, two Faraday isolators (FI) are used to prevent back propagations. An electro-optic modulator (EOM) is used to modulate the phase of the laser light. An acousto-optic modulator (AOM) is used to control the frequency and intensity of the laser. The cavity table is connected to the laser table with a single-mode optical fibre. On it most of the element of Fig.~\ref{fig1} are recalled. In addition to the $P_e$ and $P_t$ photodiodes, the $P_{inj}$ and $P_{PDH}$ ones are represented and respectively monitor the intensity injected into the cavity and the one reflected by it.}\label{fig2}
\end{figure}

The Fig.~\ref{fig2} shows a scheme of the experimental apparatus used in the present work. As already mentioned, it follows the 1979 proposal of Iacopini and Zavattini \cite{Iacopini1979} as all the reported experiments since then \cite{Battesti2013}.

\subsection{Laser source and frequency locking}

The laser source used in our experiment is a Nd:YAG Laser of 1064~nm wavelength and 1~W power.

The laser light polarisation coming out of the laser unit is rendered linear thanks to a $\frac{\lambda}{4}$ and aligned into a Faraday isolator thanks to a $\frac{\lambda}{2}$. It is then guided into an electro-optic modulator (EOM) that is used to modulate the phase of the laser light. The light continues into an acousto-optic modulator (AOM) in double pass configuration. This AOM is used to control the frequency of the laser and the beam intensity. The light is then coupled into a single-mode optical fibre with a typical transmission efficiency of 85\%. This optical fibre delivers the light to the cavity in vacuum. A part of it ($\approx$ 20\%) is redirected through a window of the vacuum chamber to the photodiode $P_{inj}$ that is used to monitor the power incident to the cavity and to stabilise it. Through the same window, light reflected by the cavity is collected by another photodiode $P_{PDH}$, see Fig.~\ref{fig2}. This intensity is necessary for the Pound-Drever-Hall method \cite{Drever1983} to lock the laser frequency to the Fabry-Perot cavity resonance frequency. 

\begin{figure}[ht!]
\centering
\includegraphics[width=1\textwidth]{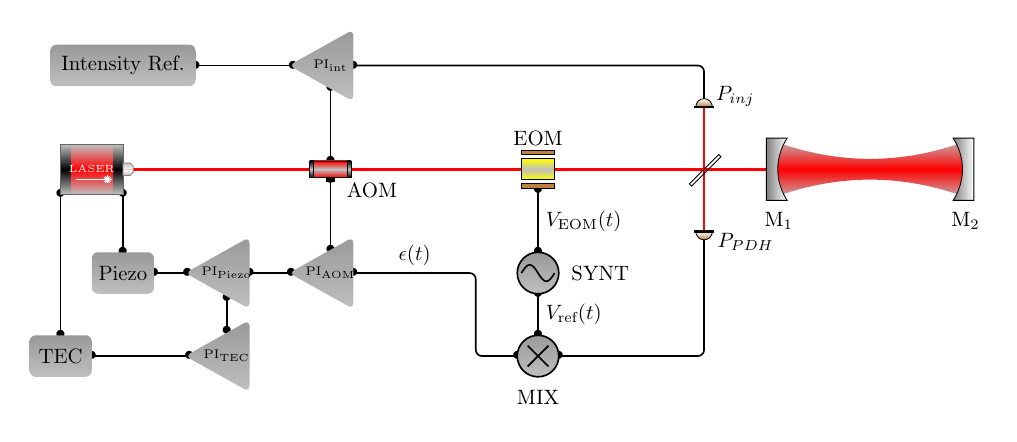}
\caption{Locking frequency and intensity system scheme used in the BMV experiment. The error signal $\epsilon (t)$ is produced by a mixer (MIX) by multiplication of $P_{PDH}$ voltage signal in the cavity reflection with $V_\text{ref}(t) = V_\text{ref}\sin{\Omega t}$ voltage which is produced by a synthesiser (SYNT) which also sends a $V_\text{EOM}(t) = V_\text{EOM}\sin{\Omega t}$ to modulate in phase the laser beam. After that, the error signal $\epsilon (t)$ is sent to 3 proportional integrator (PI) regulators in a row to do fast corrections with an acousto-optic modulator (AOM),  slow corrections with a piezoelectric controller (Piezo) and very slow corrections with a thermo-electrical Peltier module (TEC) on the laser frequency.}\label{fig5}
\end{figure}

In Fig.~\ref{fig5} we show a scheme of our laser locking system.
An error signal $\epsilon(t)$ is obtained by demodulating $P_{PDH}$ photodiode power thanks to a mixer at the driving frequency of the phase modulation. The error signal $\epsilon(t)$ is injected into three regulators in series to have a fast correction of the laser frequency (AOM), a slow one (Piezo) and a very slow one (TEC). The Piezo and TEC temperature correction, acting directly on the laser crystal, are included in the laser system itself.

\subsection{Polariser prism and quarter wave-plate}

The light provided by the laser source is polarised by a polariser prism P and analysed by a polariser prism A. Following its polarisation, the laser beam splits in an extraordinary beam and an ordinary one of powers $P_e$ and $P_t$, respectively. Ideally, when the two prisms are crossed, i.e. the transmitted polarisation by the analyser is at 90$^{\circ}$ with respect to the incoming polarisation defined by P, $P_e = 0$. Thus, any change in the polarisation of the incoming light is seen as an increase of the power $P_e$ transmitted by the analyser, A. In reality, $P_e = P_t\sigma^2$ where $\sigma^2$ is called the extinction ratio and it is linked to both polariser prisms. The extinction ratio is therefore the fundamental limit to the apparatus sensitivity.

Up to the proposal of ref.~\cite{Iacopini1979}, measurements of Cotton-Mouton effects in gases were conducted using static fields and therefore by comparing the signal obtained with the magnetic field turn on to the signal obtained with the magnetic field turn off \cite{Battesti2013}.

In this case the critical experimental parameter is the polariser extinction $\sigma^2$, since $P_e = P_t\sigma^2 + P_t \Psi^2$. In the ideal case of a shot noise limited experiment, the sensitivity $\Psi_{DC}$ is 
\begin{equation}
\Psi_{DC} =\sqrt{\sigma^2}\left(\frac{h\nu}{qP_t}\right)^{\frac{1}{4}},
\end{equation}
that for a $P_t = 0.1$~W gives $\Psi_{DC} \approx 4\cdot 10^{-5} \sqrt{\sigma^2}$. It is clear therefore that the first step towards a VMB measurement is to look for polarisers which show the best extinction factor $\sigma^2$.

The best extinction ever obtained, $\sigma^2 = 2.9\cdot 10^{-10}$, is reported in ref.~\cite{Takubo1998} using a pair of Glan-Taylor polarisers \cite{Archard1948}. These polarisers are constituted each by two calcite prisms separated by an air space. As reported in ref.~\cite{Takubo1998}, Takubo and co-authors to reach their record extinction have mounted the polariser and the analyser prisms on precise tilt stages to align appropriately the beam with respect to the prisms surfaces and moreover they have used several diaphragms to be sure that as little stray light as possible could reach the photon detector. As a matter of fact, by construction, the incident angle on the entrance surface and on exit surface of each polarizing prism is a critical parameter. Actually, any lack of perpendicularity with respect to the entrance surface may be a cause of ellipticity on the extraordinary ray as in any uniaxial anisotropic medium when the ray does not propagate perpendicular to the birefringence axis. This manifests itself as an increase of extinction $\sigma^2$. In our experiment the polariser and the analyser are of the Glan-Laser type \cite{Lambrecht}.

At the exit of the optical fibre (see Fig.~\ref{fig2}) the laser beam passes through a collimator to match the beam waist with the one of the Fabry-Perot cavity. The waist divergence at the level of the polariser and the analyser is about 3$\cdot$10$^{-2}$~degrees. As expected, we have measured polariser transmissions of about 98\%. As shown in Fig.~\ref{fig2}, before the photodiode measuring the extraordinary ray power, we have a pair of diaphragms to stop as much stray light as possible. We set their diameter typically at 7~mm each. This value has been determined by studying the beam losses as a function of diaphragm diameter value.

\subsection{Optical cavity}
 
To increase the interaction time between the polarised light and the magnetic field $B$, light is confined in a region delimited by two interferential mirrors (see e.g.~\cite{Bielsa2009}), $M_1$ and $M_2$, which constitutes an optical resonant cavity of the Fabry-Perot (FP) type (see e.g.~\cite{Hartman2017} and \cite{Fabry1899}).
To ensure a maximum of cavity transmission, the laser frequency has to be kept equal to the resonance frequency of the cavity. This is obtained using the standard Pound-Drever-Hall (PDH) method (see e.g.~\cite{Berceau2010} and references within). 

The power $P_t$ transmitted at resonance by the cavity depends on the reflectivity and the losses of the mirrors. Thanks to the cavity, the ellipticity effect to be measured is multiplied by a factor $G=\frac{2 F}{\pi} $ \cite{Hartman2017}, where $F$ is the cavity finesse which depends on reflectivity of both constituting mirrors \cite{BornWolf}. Interferential mirrors are known to be birefringent (see e.g.~\cite{AgilLetourneur2023} and references within) thus FP cavity also induces an ellipticity $\Gamma$ on the laser beam going through it. Thus, if a birefringent medium is in between the polariser P and the analyser A, so that light acquires an ellipticity $\Psi$ in particular going through the optical cavity, $P_e$ can be written as 
\begin{equation}
\label{Pext}
P_e = P_t\sigma^2 + P_t\left(\Gamma + \Psi\right)^2 = P_{DC} + 2P_t\Gamma\Psi + \Psi^2P_t \quad \quad >0.  
\end{equation}
One obtains $\Psi + \Gamma$ as
\begin{equation}
\Psi + \Gamma = \pm\sqrt{\frac{P_e}{P_t}-\sigma^2},
\end{equation}
where 
\begin{equation}
\Gamma = \pm\sqrt{\frac{P_{DC}}{P_t}-\sigma^2}.
\end{equation}
In our case, since $\Psi \ll \Gamma$, $\Psi$ can be approximated as
\begin{equation}\label{Psieq}
\Psi = \frac{P_e - P_{DC}}{2P_t\Gamma} =  \frac{\Delta P_e}{2P_t\Gamma}.
\end{equation}
The sign of $\Gamma$ is usually retrieved thanks to a calibration by a measure of a known linear magnetic birefringence of a gas, like e.g. molecular nitrogen. Once the sign of $\Gamma$ is known, since the sign of $\Delta P_e$ is also known, the sign of $\Psi$ can be established.
The ultimate limitation of such a measurement is given by the shot noise related to the detection of the $P_e$ power. Following \cite{Battesti2008}, the corresponding signal to noise ratio $r$ can be written as 

\begin{equation}
r \approx \Psi\sqrt{\frac{4qP_t\Gamma^2}{\left(\sigma^2 + \Gamma^2 \right)h\nu}},
\end{equation}
where $q$ is the quantum efficiency of the detector, and by definition the acquisition time is assumed to be 1 second. $r$ is therefore given in rad/$\sqrt{\text{Hz}}$. If one calls sensitivity the ellipticity $\Psi_S$ which gives $r = 1$, one obtains
\begin{equation}
\Psi_S \approx \sqrt{\frac{h\nu}{4qP_t}\left(1+\frac{\sigma^2}{\Gamma^2}\right)},
\end{equation}\label{Bestsens}
which seems to indicate that the best is to have $\Gamma^2 \gg \sigma^2$.

Our FP optical cavity is constituted by two commercial spherical interferential mirrors (see e.g.~\cite{Bielsa2009}) of intensity reflectivity $R$, intensity transmission $T$, intensity losses $P$, with, by definition, $1 = R + T + P$, and a curvature radius $R_C=2$~m. For the sake of simplicity, in the following we will assume that the two mirrors were identical. At our wavelength of 1064~nm, the mirror intensity reflectivity that has been reported is very close to 1, $R=$~0.99999601~$\pm$~0.00000006 \cite{AgilLetourneur2023} and therefore $T,P \sim 2$~ppm.

The cavity is aligned so that only the TEM$_{00}$ mode resonates \cite{Kogelnik1966}. The cavity resonance linewidth $\Delta\nu_c$ can be written as $\Delta\nu_c = \frac{c}{2L_cF}$, where $F$ is the cavity finesse, $F = \frac{\pi \sqrt{R}}{1-R}$. The highest finesse ever reported at $\lambda=$1064~nm is 895~000 \cite{AgilLetourneur2023}. 

The cavity finesse is usually derived by the measurement of the cavity lifetime $\tau_c$ which is the characteristic $\frac{1}{e}$ time of the exponential decay observed by the transmitted power when the incident light of the cavity is suddenly turned off, $\tau_c = \frac{L_c F}{\pi c}$ \cite{Berceau2012}. In our case, $\tau_c$ is around 1~ms. The best $\tau_c$ ever reported at $\lambda=$1064~nm is about 3.6~ms obtained by a 123~m long FP cavity of 27~500 finesse \cite{kozlowski2025design}, while the best $\left(\frac{1}{e}\right)$ $\tau_c$ reported in an experiment similar to ours, for a 3.3~m long FP cavity, at 1064~nm and a finesse 770~000, is 2.7~ms \cite{DellaValle2014}.

When the cavity is at resonance, the ratio $T_C$ between the transmitted power $P_t$ exiting the cavity and the injected one $P_i$ entering into it can be written as $T_C = \frac{P_t}{P_i} = \left(\frac{T}{T + P}\right)^2$. This ratio tends to 1 only in the case that $T \gg P$, but when $R \approx 1$, as in our case, assuming $T \approx P$, $T_C \approx 0.25$. All this applies only if a single mode resonates and all losses not depending on mirrors themselves are negligible. During operation our $T_C$ always exceeds a few percent. 

\subsubsection{Cavity as a low-pass filter}

As detailed in ref.~\cite{Agil2021} and refs. within, a FP cavity has a different dynamical behaviour in the ordinary beam than in the extraordinary one. $P_t$ time response is a typical first-order low-pass filter, while the extraordinary beam $P_e$ reacts as if the cavity had filtered it twice. The cavity filter bandwidth can be written as $\nu_{bw} = \frac{c}{4L_cF}$. During our run 2021 and run 2025 our FP cavity had a $\nu_{bw}$ between 50 and 70~Hz.

\begin{figure}[h!]
    \centering
    \includegraphics[width=\textwidth]{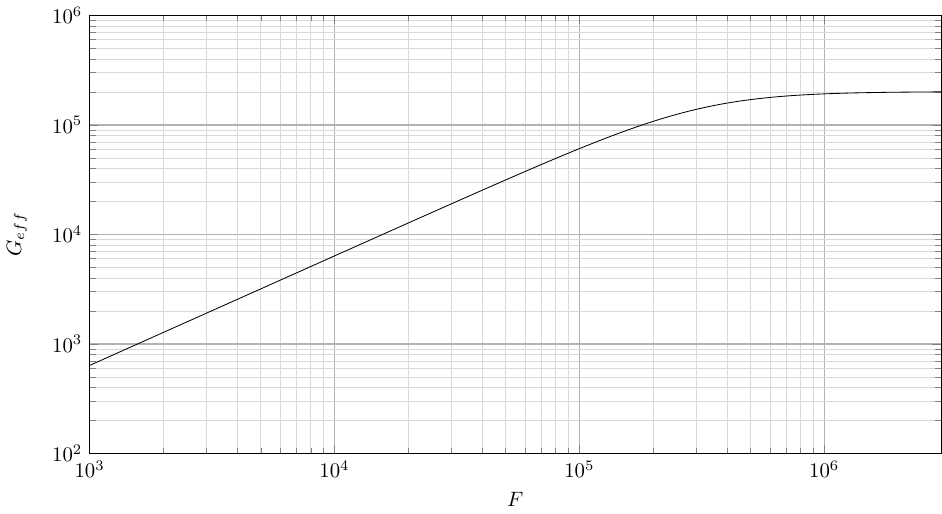}
    \caption{$G_{eff}$ as a function of $F$, with $L_c = 2.55$ m and $\nu_p \approx 100$ Hz.}
    \label{fig10}
\end{figure}

Now, assuming that only the ellipticity $\Psi$ varies in time and not any other experimental parameter, as again shown in \cite{Agil2021}, $P_e$ depends on the value of $B^2_f(t)$ i.e. the square of the magnetic field pulse filtered by the cavity instead of the original $B^2(t)$ one. This means that increasing the finesse by a factor does not correspond necessarily to an increase of the same factor in the ellipticity $\Psi$ to be measured because the maximum of the filtered square of the magnetic field decreases when the gain factor $G$ increases \cite{Agil2021}. At the limit when $F$ is such that $\nu_{bw} \ll \nu_p$, where $\nu_p$ is the main harmonic frequency of $B^2(t)$, the effect to be measured does not increase any more increasing $F$ since, 

\begin{equation}
\Psi \propto G_{eff} = \frac{2F}{\pi\sqrt{1+\left(\frac{\nu_p}{\nu_{bw}}\right)^2}}= \frac{2F}{\pi\sqrt{1+F^2\left(\frac{\nu_p 4L_c}{c}\right)^2}},
\end{equation}
as shown in Fig.~\ref{fig10}. In our case $G_{eff} \approx \frac{2F}{\pi\sqrt{1+F^2 10^{-11}}} \approx 0.6\frac{2F}{\pi}$.

\subsubsection{Cavity as a wave-plate}
\begin{figure}[ht!]
\centering
\includegraphics[width=1\textwidth]{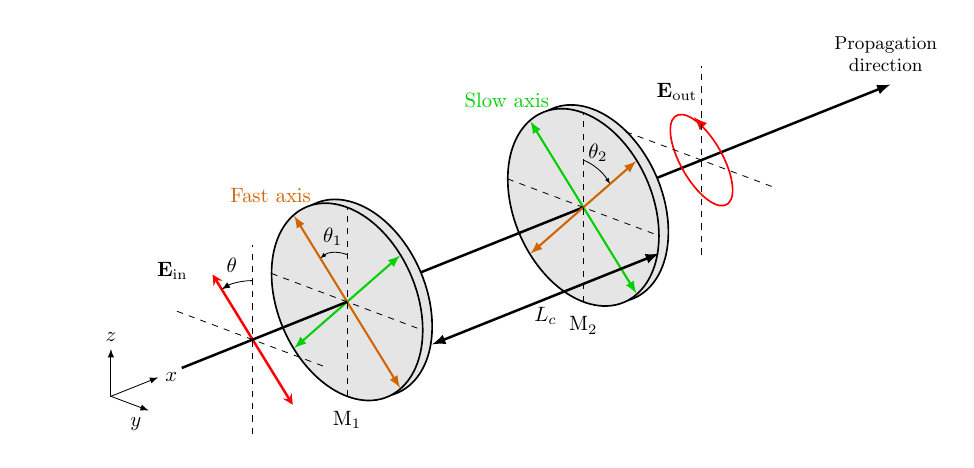}
\caption{Schematic representation of a birefringent cavity. Each mirror M$_i$ has a phase shift $\Gamma_i$ between its fast and slow axes and is rotated by an angle $\theta_i$ relative to a reference. Depending on the angle of incident polarisation $\theta$ with respect to the same reference for $\theta_i$, the acquired ellipticity is equal to $\Gamma = \Gamma_1 + \Gamma_2$ which is given by eq.~(\ref{eq:Gamma}).}\label{fig11}
\end{figure}

As discussed in ref.~\citep{AgilLetourneur2023} and references within, interferential mirrors act as wave-plates. Each mirror has its own birefringence axis whose origin is not clearly understood yet \citep{AgilLetourneur2023}. The optical cavity can be therefore modelled as a combination of two wave-plates (see Fig.~\ref{fig11}) that originate a total ellipticity $\Gamma$ that can be written as
\begin{equation} \label{eq:Gamma}
\Gamma = \Gamma_1 + \Gamma_2 = \frac{2F}{\pi}\left(\frac{\delta_1}{2}\sin{\left(2(\theta_1 - \theta)\right)} + \frac{\delta_2}{2}\sin{\left(2(\theta_2 - \theta)\right)}\right),
\end{equation}
where $\delta_{i=1,2} \ll 1$ rad, is the phase shift accumulated between the slow and fast axis with a single pass through the mirror M$_i$. Thus, $\Gamma$ can be changed by appropriately rotating both mirrors i.e. changing $\theta_1$ and $\theta_2$, going from zero to a maximum value depending on the intrinsic birefringence of each mirror \cite{AgilLetourneur2023}. 

Once the zero $\Gamma$ is obtained any small value of $\Gamma$ can be obtained by a small rotation of one of the two cavity mirrors. Obviously, all this procedure has to be done while the laser frequency is locked to the cavity. The condition $\Gamma = 0$, with $\delta_1 \neq \delta_2$ also means that the birefringence axis of the cavity is aligned with respect to the light polarisation.

The value of birefringence and the axis orientation seem to depend on the point of reflection of light on the mirror reflecting surface as well. Moreover, this dependence does not seem to be due to a random process \cite{Micossi1993}. Thus any movement of the mirror with respect to the incoming light and vice versa changes the total ellipticity acquired by the light while trapped in the optical cavity since it changes the optical axis of the cavity and the light impact points on the mirrors \cite{Solimeno1991}. The measurements reported in ref.~\cite{Hartman2019} indicate that a tilt of about one $\mu$rad of one mirror may originate a change of ellipticity of the order of 10$^{-4}$. This is likely the explanation why one can decrease the extinction ratio by tilting mirrors as we reported in subsection~\ref{exc}.

The fact that the FP cavity acts as a birefringent medium also implies the existence of a frequency splitting between $P_t$ and $P_e$ components because of their respective polarisations which are at 90$^\circ$ to each other.

As discussed in \cite{Berceau2012}, this can create an accuracy error of the order of less than 1\%, which is negligible at the present stage. This applies when the $P_t$ component is locked at the top of the FP Airy transmission function \cite{Berceau2012}. If locking conditions change in time the ratio $\frac{P_e}{P_t}$ changes as well giving a contribution to the apparent birefringence noise, as discussed in the following.
\subsection{Foil Coil}

\begin{figure}[ht!]
\centering
\includegraphics[width=1\textwidth]{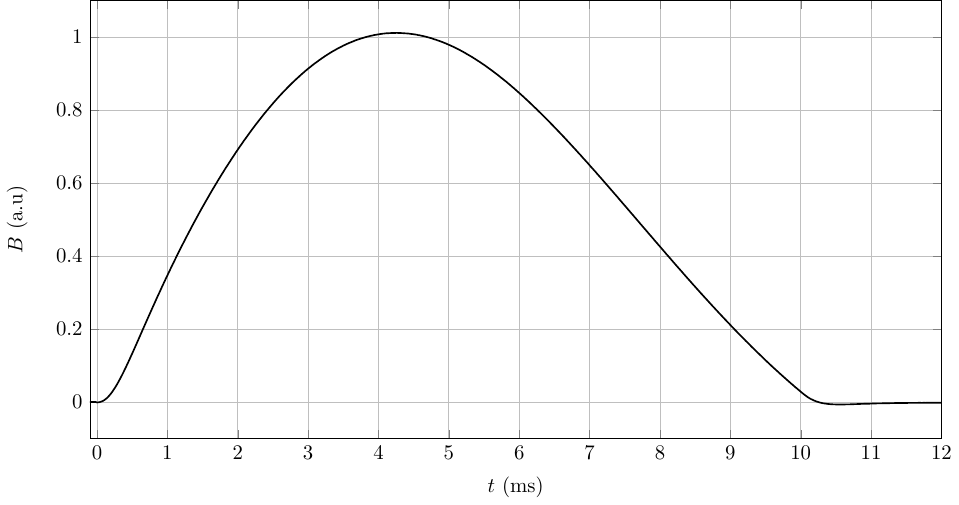}
\caption{Foil Coil magnetic field $B(t)$ profile with a rising time of about 4.3~ms and a total duration $T_D$ of about 10~ms.}\label{fig3}
\end{figure}

An external magnetic field $B$ can be generated in the region between the two cavity mirrors by the use of pulsed coils which is the particularity of our experiment \cite{Beard2021}. Fig.~\ref{fig3} shows a typical magnetic pulse produced by our Foil Coils. The pulse duration $T_D$ is a few milliseconds. The mathematical form is similar to a sine function $B(t) = B_{max}\sin(2\pi\nu_p t)$ for almost the whole pulse duration.

Actually, pulsed fields are generated by discharging a bank of capacitors in a coil. At the beginning, each pulse is described by the oscillating solution of a typical $R_t, L_t, C_t$ electric circuit, where $R_t$, $L_t$ and $C_t$ are respectively the total resistance, inductance and capacitance of the circuit. When the driving voltage $U$ changes sign, the appropriate diodes prevent the capacitors bank from being almost fully recharged and the circuit current decreases exponentially as a typical $LR$ circuit. $t_{max}$, the time necessary to get the maximum field $B_{max}$ is proportional to $\sqrt{L_tC_t}$. For example, for the pulse shown in Fig.~\ref{fig3} the driving voltage changes sign at about 5~ms. 

Typically, we drive our pulsed magnet thanks to a 3~MJ bank of 40 capacitors obtaining a pulse about 14~ms long. Pulse duration can be reduced to about 10~ms reducing the number of capacitors to 20 (see Fig.~\ref{fig3}) and to about 6~ms using only 10 capacitors. To obtain a 10~T field on a meter length and 0.01~m radius section, one needs to discharge more than 100~kJ in a few milliseconds corresponding to about 100~MW peak power.

Since the expected vacuum magnetic birefringence is proportional to $B^2$ (see formula eq.~(\ref{VCM})), finally the ellipticity to be measured is
\begin{equation}
\label{PsiG}
\Psi(t) = G\times\pi\frac{L_B}{\lambda}\Delta n_{VMB} = 2F\frac{L_B}{\lambda}K_{VMB}B(t)^2.
\end{equation}
For the sake of the argument, if one assumes that $F=10^6$, $L_B=1$~m, $\lambda=10^{-6}$~m, $B(t)=10$~T, $\Psi \approx 8\cdot 10^{-10}$~rad.

To increase the signal to noise ratio, one typically acquires the signals $P_t$ and $P_e$ for $n$ identical pulses obtaining $n$ different $\Psi(t)$. The final value of $K_{VMB}$ is deduced by the average of all the $\Psi(t)$. 

The technical details on the Foil Coil, the pulsed magnet specially designed for the BMV project and used for the runs 2021 and 2025, have already been reported in ref.~\cite{Beard2021} on the occasion of its commissioning. 

\begin{figure}[h!]
\centering
\includegraphics[width=0.5\textwidth]{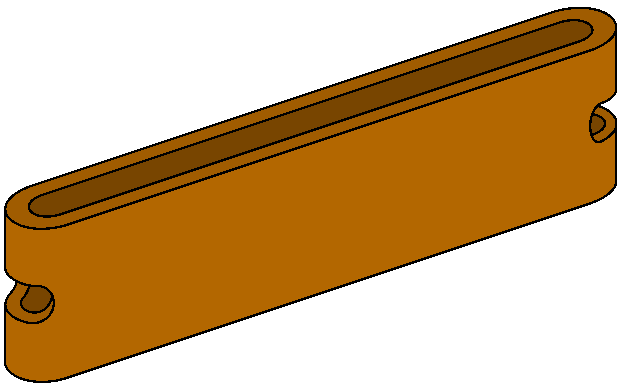}
\caption{Schematic view of the conductor shape once wound.}\label{fig13}
\end{figure}

We just recall that the magnet is based on the winding of a copper foil insulated with two layers of Kapton tape. About 100 m of the copper foil is wound over a glass fibre epoxy FR4-type body with a racetrack shape, representing about 50 layers of conductor. The optical access is provided by two holes in each turn in the insulated copper (see Fig.~\ref{fig13}).

\begin{figure}[h!]
\centering
\includegraphics[width=0.5\textwidth]{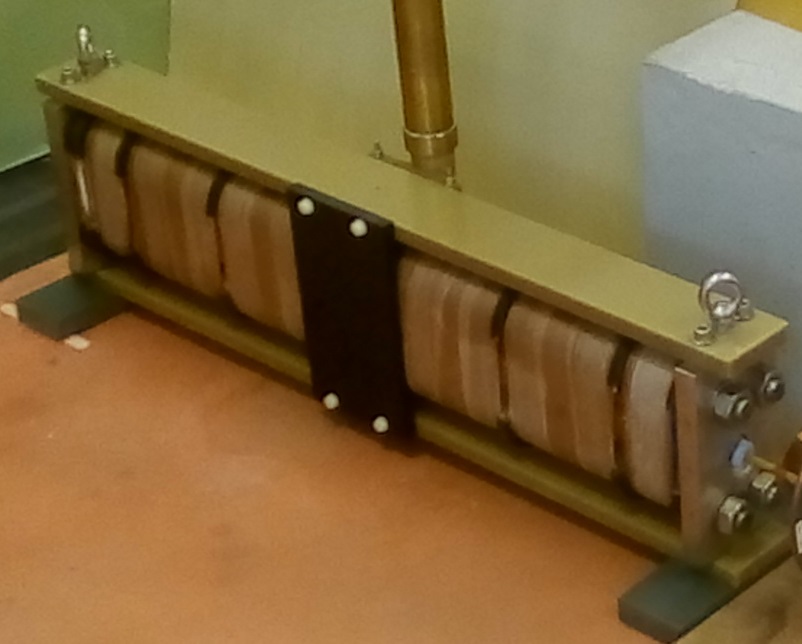}
\caption{Photo of the Foil Coil used in the run 2021.}\label{fig14}
\end{figure}

Once the coil is appropriately reinforced it appears as shown in Fig.~\ref{fig14}. This coil has been tested up to 12.5~T \cite{Beard2021}.

\subsection{Data analysis}\label{subsec37}

Our time data analysis follows what has been detailed in ref.~\cite{Cadene2014}. According to eq.~(\ref{Psieq}), assuming $\Psi \ll \Gamma$, the signal to  be measured to derive $\Psi(t)$, $\Delta P_e$, is odd with respect in the transformation of $\Gamma$ into $-\Gamma$ and even with respect to the transformation of $B$ into $-B$. If we take four series of data with ($\Gamma>0$, $B>0$), ($\Gamma<0$, $B>0$), ($\Gamma>0$, $B<0$), ($\Gamma<0$, $B<0$), having divided any $\Delta P_e$ by its corresponding $2P_t\Gamma$, we can build four functions averaging the $\Psi$ of the pulses constituting the series and calculate the average measured ellipticity at the instant $t$, for example $\Psi^{(>,>)}(t)$. The corresponding uncertainty at $t$ is the standard deviation of the average value obtained at the same instant. Thus 
\begin{equation}
\label{eq:TruePsi}
\Psi^{(-+)}(t) = \frac{1}{4} \left(\Psi^{(>,>)}(t) + \Psi^{(<,>)}(t) + \Psi^{(>, <)}(t) + \Psi^{(<,<)}(t)\right),
\end{equation}
where, $(>,>)$ means $\Gamma>0$, $B>0$ and so on. 

The linear combination of the four functions so worked out allows to isolate the signal with the demanded symmetries erasing any other spurious signal, in principle. To finally obtain the $K_{VMB}$ value we fit $\Psi(t)$ by the $B_f^2(t)$ function filtered by the cavity low-pass filter with $K_{VMB}$ as free parameter according to eq.~(\ref{PsiG}). 

\subsection{Noise and systematic effects}

As shown in ref.~\cite{Agil2021}, the sources of noise differ before and after the beginning of the magnetic pulse. This applies also to the possible systematic effects.

Thus, the acquisition span can be divided in different durations corresponding to the different noise sources or experimental problems which have all to be brought under control to reach the ultimate sensitivity which would allow to measure the VMB, averaging the data of a reasonable number of magnetic pulses. If, by definition, we assume that the magnetic pulse starts at the instant $t=0$, studying our pulsed signal we have to consider the following durations:
\begin{itemize}
\item[-] $t<0$: before the pulse the noise is only due to the presence of the laser light and the optical apparatus characteristics. This is what we call the optical noise. This is our best expected noise since, at the best, we can only succeed in that the magnetic field pulse does not affect the optical noise for $t>0$.
\item[-] $0<t<t_a$: This duration corresponds to the time span in which no increase of the optical noise is detected after the beginning of the magnetic pulse.
\item[-] $0<t<t_{\Psi}$: The time $t_{\Psi}$ is the time at which the signal analysis has to be stopped and therefore we have a value of $\Psi(t)$ only for $0<t<t_{\Psi}$. Obviously, it is more profitable when $t_{\Psi} \geq t_{max,B_f^2}$, where $t_{max,B_f^2}$ is the time at which the square of the B field filtered by the cavity is at its maximum value.
\end{itemize}

\subsubsection{$t<0$: Optical noise}

The ellipticity optical noise $\delta\Psi$ of our experiment has been studied in detail in ref.~\cite{Hartman2017,Agil2022}. It can be written as $\delta\Psi=\delta P_e/2P_t\Gamma$, where for any variable $v$, its noise is indicated as $\delta v$ and therefore $\delta P_e$ is the noise associated to the extraordinary beam power $P_e$.

In the general case, the expression for the ellipticity noise $\delta\Psi$ is constituted by noise associated to the experimental parameters on which $P_e$ depends, plus $P_e$ shot noise, as well as  $\delta P_{o}$ which represents any noise independent on the optical parameters
\begin{equation}
\delta P_e=\delta P_{DC}+\delta P_{shot}+\delta P_o,
\end{equation}
with $\delta P_{shot}$ the shot noise and $\delta P_{DC}$ is
\begin{equation}
\delta P_{DC}=\delta P_t\left(\sigma^{2}+\Gamma^{2}\right)+P_t\left(\delta \sigma^{2}+2\Gamma\delta\Gamma\right).
\end{equation}
Thus, we find the ellipticity noise to be
\begin{align}\label{DeltaPsiPolHo}
\delta\Psi  \nonumber  ={}& \frac{\delta P_t}{2P_t} \left( \frac{\sigma^2}{\Gamma} + \Gamma \right) + \frac{\delta\sigma^2}{2\Gamma} \\
 &+ \delta\Gamma + \frac{\delta P_{shot}}{2P_t\Gamma} + \frac{\delta P_o}{2P_t\Gamma},
\end{align}
where $\delta P_t \left( \sigma^2/\Gamma + \Gamma \right)/2P_t$ is an apparent ellipticity noise due to power fluctuations, $\delta\sigma^2/(2\Gamma)$ is an apparent ellipticity noise due to noise in polariser extinction, $\delta\Gamma$ is the true ellipticity noise, $\delta P_{shot}/(2P_{t}\Gamma)$ is related to power shot noise and $\delta P_o/(2P_t\Gamma)$ include all noises induced by causes independent on the optical parameters like the noise equivalent power (NEP) of the detection photodiode. Let's note that $\delta\Gamma$ does not depend on $\Gamma$ at the lowest order in $\Gamma$, as shown in ref.~\cite{Agil2022}.

According to eq.~(\ref{DeltaPsiPolHo}), $\delta\Psi = \alpha \Gamma + \beta + \frac{\varepsilon}{\Gamma}$ whose minimum value, $\delta\Psi_{min}$ can be reached if $\Gamma_{min} = \sqrt{\frac{\varepsilon}{\alpha}}$,  $\delta\Psi_{min} = 2\sqrt{\varepsilon\alpha} + \beta$. If, as usual, $\varepsilon \sim \frac{\delta P_t}{2P_t}\sigma^2$ and $\alpha \sim \frac{\delta P_t}{2P_t}$, $\Gamma_{min}^2 = \sigma^2$ and $\delta\Psi_{min} = \frac{\delta P_t}{P_t}\sqrt{\sigma^2} + \delta\Gamma$. Thus, one has to have the smallest value of $\sigma^2$ but then the experiment sensitivity is anyway limited by the true ellipticity noise.

\subsubsection{$0<t<t_a$: silent duration}
At $t=0$ the magnetic pulse starts. In principle, one might expect that systematic effects induced by the field or its derivative or the product of the two start appearing also. Let's note that field derivative has its maximum value at $t=0$. 

On the other hand, $t=0$ can also be the beginning of any acoustic noise i.e. any perturbation propagating in the air or in the apparatus constituents, in particular the optical table and the vacuum chambers, bellows and pipes, and reaching the optical elements in their mounts. This acoustic noise can be generated by mechanical stresses in the coil due to the derivative of the driving current. 

Acoustic noise obviously takes more time to arrive from its source the coil to optics than its electromagnetic counterpart that is practically instantaneous. 

As already said, what we call the silent duration is the span during which no increase of the optical noise is detected.

Following \cite{Agil2021}, in our 2014 data \cite{Cadene2014} $t_a \approx 0.150$~ms while in run 2021 and run 2025 $t_a$ exceeds 1~ms.

\subsubsection{$0<t<t_\Psi$: analysis duration} 

At the end of the silent duration discussed in the previous paragraph the statistical noise increases and systematic effects mimic a non-zero signal \cite{Agil2021}. It becomes clear therefore that signal cannot any more be modelled as a function proportional to the square of $B(t)$ filtered by the cavity. In 2014 analysis has been stopped at 3.1~ms since an important perturbation of the locking error signal was observed at 3.2~ms \cite{Cadene2014}, even if $t_a$ was about 0.1~ms. In the present analysis of the whole data of run 2021 and run 2025, we stop the analysis following some criteria detailed in the next section. The values of $t_\Psi$ used are reported in results tables \ref{tab4} and \ref{tab6}. Let's note $t_\Psi \approx t_a$.

\subsubsection{Systematic effects}
\label{Systeff}
The whole data analysis, starting from eq.~(\ref{Pext}), is based on the assumption that only $\Psi(t)$ varies in time, while all other experimental parameters are constant. This is obviously an ideal point of view. Actually, all the parameters may vary in time, in particular for $t>0$ when the perturbation associated to the magnetic pulse propagates in the apparatus.

For the sake of the argument, let's assume that $\Psi(t)=0$ and that for each variable $v$

\begin{equation}
v(t) = v(t\leq 0) + \Delta v(t),
\end{equation}

where $v(t\leq 0)$ corresponds to a constant value and $\Delta v(t)$ is the variation of $v(t)$ during the pulse, $\Delta v(t) \ll v(t \leq 0)$. For the sake of simplicity, let's also assume that $v(t)$ varies slowly so that we can neglect the cavity filtering.

Following eq.~(\ref{Pext}), a spurious ellipticity signal $\Psi_{\text{sp}}$ appears mimicking a true ellipticity signal

\begin{align}
2P_t(t\leq 0)\Gamma(t\leq 0)\Psi_{\text{sp}}(t)\approx & P_t(t\leq 0)\Delta \sigma^2(t) \nonumber \\
& + \Delta P_t(t)\sigma^2(t\leq 0) \\
& + 2P_t(t\leq 0)\Gamma(t\leq 0)\Delta \Gamma(t) \nonumber \\
& + \Delta P_t(t)\Gamma^2(t\leq 0). \nonumber 
\end{align}

Now, thanks to $\left(\Gamma,B\right)$ symmetry analysis described in the subsection \ref{subsec37}, since $\Delta\sigma^2(t), \Delta P_t(t), \Delta \Gamma(t)$ do not depend on the sign of $\Gamma$, the only term remaining in principle is the one containing $\Gamma(t\leq0)$ and therefore

\begin{equation}
\Psi_{\text{sp}}^{(-+)} = \Delta\Gamma(t).
\end{equation}

\section{Run 2021 \& Run 2025}
\label{sec3}
\subsection{Generalities}
During and since the commissioning of the novel pulsed magnet reported in \cite{Beard2021}, we have performed thousands of pulses. Let's note that Foil Coil technology allows us to perform a pulse duty cycle up to 10 pulses at 5~T per working hour. Currently, the main limitation to increase the pulse duty cycle is the time spent between each pulse to realign optics if necessary and check data to be sure that there is no sign of damage on the coil. Most of the pulses have been used to study noises and the acoustical perturbation propagation induced by the pulse itself. Some of them have been performed in the presence of gases like helium or nitrogen. Comparing the result obtained for a gas effect to the one expected by theory or to the other existing experimental values is a standard way to test the accuracy of a VMB apparatus (see e.g.~\cite{Ejlli2020}). About a thousand of them were performed in high vacuum with pressure ranging from about 10$^{-6}$ to 10$^{-7}$~mbar thanks to two ionic pumps. Following \cite{Rizzo1997}, the Cotton-Mouton effect of residual gases is negligible with respect to the expected sensitivity.

Between run 2021 and 2025 a certain number of changes to the basic apparatus have been put in place to improve the performances of our setup. They are discussed in the following.

\subsubsection{Extinction ratio $\sigma^2$}

\label{exc}

During run 2021, in the presence of the cavity, we have reached extinctions as low as a few 10$^{-9}$ \cite{Beard2021} by carefully aligning the optical beam in particular with respect to the cavity mirrors. Actually, any tilt of one of the cavity mirrors changes the cavity optical axis \cite{Solimeno1991} changing therefore the incidence angle on the analyser prism and the output beam point on the opposite mirror. An explanation why this can change the extinction has been given in the previous section where we deal with the optical cavity. We have thus optimized both the cavity mode and the extinction.

During run 2025, we have simplified the procedure by introducing between the polariser and the analyser a $\lambda$/4 wave-plate. The idea is that any lack of alignment of the polariser prism creates an ellipticity that can be changed into a rotation angle by the use of the $\lambda$/4 wave-plate and compensated by rotating the analyser. By this technique we have reached extinction ratio better than 10$^{-8}$ without the cavity which is a net improvement with respect to the $10^{-7}$ obtained before, since now the alignment of the cavity is decoupled of the alignment of the polarisers for the maximum extinction. 

\begin{figure}[ht!]
\centering
\includegraphics[width=1\textwidth]{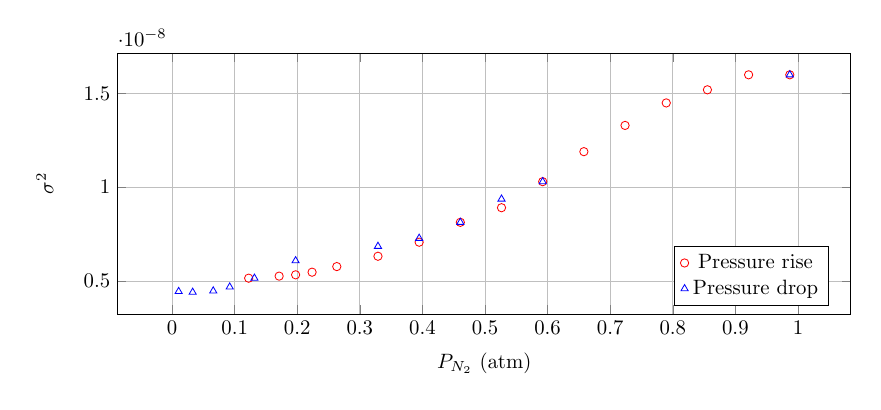}
\caption{Variation of the extinction ratio $\sigma^2$ with the pressure $P_{N_2}$ in gaseous nitrogen. The minimum of $\sigma^2$ is adjusted  at the lowest pressure.}\label{fig8}
\end{figure}

Let's note that the polariser and analyser have to be aligned in vacuum to reach the best extinction ratio since this one changes going from 1~atm of N$_2$, for example, to high vacuum and vice-versa as shown in Fig.~\ref{fig8}. 

The typical $\sigma^2$ values for each set are given in tables \ref{tab3} and \ref{tab5}.

\subsubsection{Cavity locking}

\begin{figure}[ht!]
\centering
\includegraphics[width=1\textwidth]{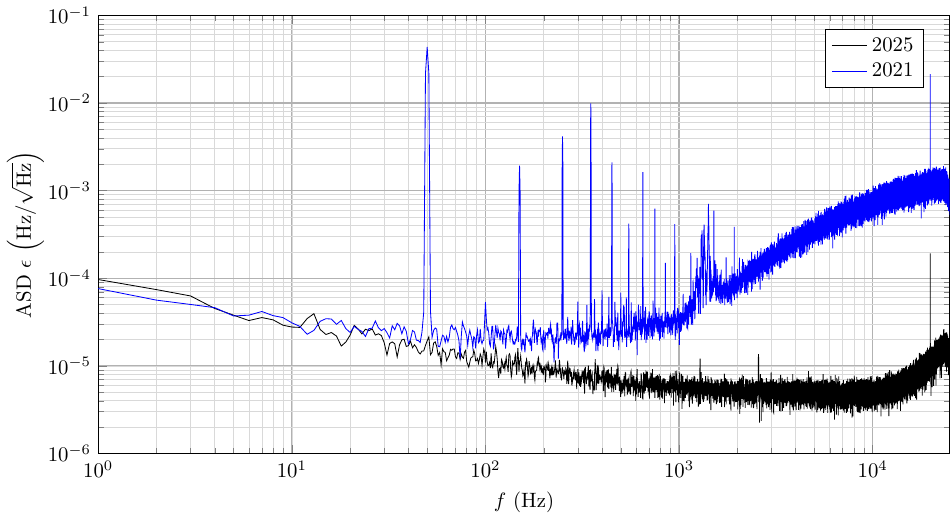}
\caption{ASD of the error signal $\epsilon$ as a function of frequency of the run 2021 (blue) and 2025 (black) \cite{agil2023birefringence}.}
\label{fig6}
\end{figure}

The Fig. \ref{fig6} shows the Amplitude Spectral Density (ASD) of a typical closed-loop error signal $\epsilon(t)$ for run 2021 and 2025 where an improved locking system has been used. These more recent opto-electronic devices allow us to have a bigger gain which gave a larger bandwidth and killed most of the resonances of the spectrum. For this ASD and for all the following, we obtain ASD by using the square root of the Power Spectral Density (PSD) with Welch's method thanks to the SciPy library \cite{2020SciPy-NMeth}.

The minimum value of ASD shown in Fig. \ref{fig6} divided by the cavity linewidth corresponds to about $10^{-6}~\text{Hz}^{-\frac{1}{2}}$ which is a standard result, for example already obtained in 1996 \cite{de1996very}.

\subsubsection{Off-cavity wave-plate}
The role of $\Gamma$ can be played by a wave-plate situated before the cavity mirrors as shown in Fig.~\ref{fig2}. The advantage of this is to decouple the establishment of the D.C. carrier ellipticity that beats with the modulated one induced by the magnetic field (see eq.~(\ref{Pext})) from the alignment of the cavity which is always somewhat critical. 

The ellipticity $\eta$ induced by this off-cavity wave-plate has to be as small as cavity $\Gamma$ without deteriorating the polariser extinction. We therefore took a non-birefringent optical plate of a quality that does not interfere with extinction and we induced on it a lateral mechanical stress inducing some stress birefringence. When this stress is low enough and $\eta$ consequently, extinction is not affected by the presence of this off-cavity wave-plate. We introduced this $\eta$ ellipticity during run 2025.

The ellipticity induced by this off-cavity wave-plate is such that the major axis of its polarisation is at the first order aligned with the incoming light polarisation \cite{BornWolf}.

With the $\eta$ wave-plate in place we set $\Gamma$ as close as possible to 0. Thus, the eq. (\ref{eq:TruePsi}) still applies by replacing $\Gamma$ by $\Gamma + \eta$ and deal with the symmetries of $\eta$ instead of $\Gamma$. 

\begin{figure}[h!]
\centering
\includegraphics[width=1\textwidth]{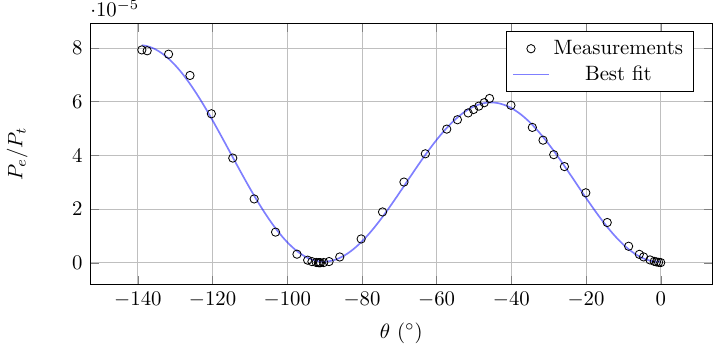}
\caption{The ratio $P_e/P_t$ as a function of the rotation angle $\theta$ of the wave-plate $\eta$ relative to an arbitrary fixed reference angle $\theta_0$. A best fit using the function of $\theta$: $\left[A\sin{\left(2(\theta - \theta_0)\right)} + B\right]^2 + C$ is superimposed to the experimental data ($A = \left(3.34 \pm 0.03\right) \cdot 10^{-3}$~rad, $B = \left(6.4 \pm 0.3\right) \cdot 10^{-4}$~rad, $C = \left(4.3\pm 2.7\right) \cdot 10^{-7}$~rad$^2$, $\theta_0 =(88.4\pm0.3)^\circ$).}\label{fig12}
\end{figure}

Another advantage of this way to introduce a D.C. ellipticity is that one can calibrate the wave-plate once and for all and therefore knowing the value of $\eta$ by knowing the angle between the wave-plate axis and the light polarisation. In Fig.~\ref{fig12} we report typical data obtained by turning the $\eta$ in between our two crossed polarisers.
To fit the experimental data of Fig.~\ref{fig12} one needs to add another ellipticity $\eta_\text{r}$ that therefore can be acquired by light only by the passage through the analyser prism. This confirms that polariser prism may induce a small ellipticity and confirms the utility of the use of quarter-wave plate as we did during run 2025.

\subsubsection{Optical noise}
\begin{figure}[ht!]
\centering
\includegraphics[width=\textwidth]{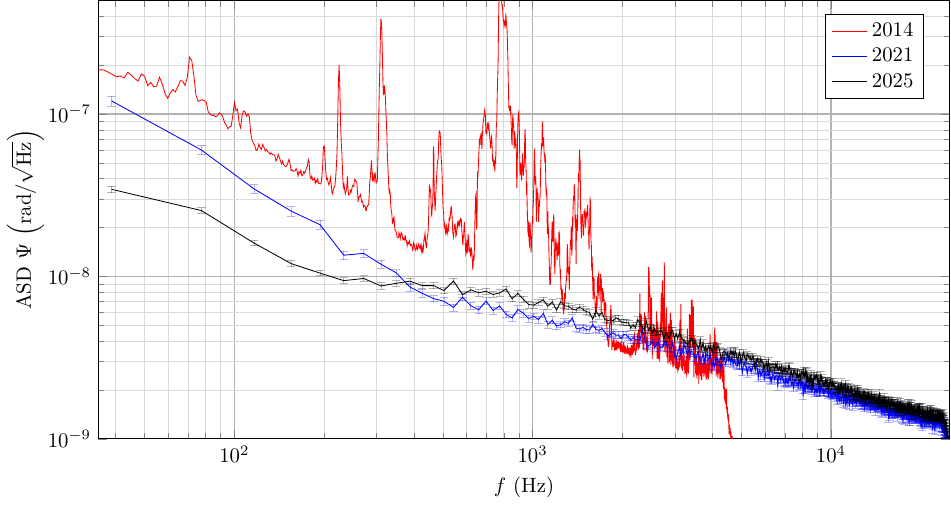}
\caption{ASD of $\Psi(t)$ before the beginning of the magnetic pulse ($t<0$), measured during run 2021, 2025 and the one reported in \cite{Cadene2014}.}\label{fig18}
\end{figure}

In Fig.~\ref{fig18} we show the average ASDs of the ellipticity $\Psi(t<0)$ corresponding to our run 2021 and run 2025 compared to the previous one of 2014 \cite{Cadene2014}.

For frequencies less than 400 Hz, Fig. \ref{fig18} shows an improvement in 2025 with respect to 2021 by a small factor (e.g. 3 at 40 Hz). As shown by us \cite{Agil2022}, the limiting noise is the one of the intrinsic cavity birefringence $\Gamma$, $\delta \Gamma$ (see eq. (\ref{DeltaPsiPolHo})). As far as we understand, this noise is induced by a lack in pointing beam stability provided by the fact that the birefringence depends on the impact point on the mirror surface.

This improvement can only be explained by changes related to the cavity done between 2021 and 2025. These are essentially the use of the off-cavity wave-plate $\eta$, that allows us to set $\Gamma$ to 0 during a set of measurements, working therefore in a stabler optical configuration. Actually, reducing $\sigma^2$ and consequently $\Gamma$ by tilting cavity mirrors gave an optical configuration very sensitive to any perturbation.

\subsubsection{Foil Coil}

\begin{figure}[h!]
\centering
\includegraphics[width=\textwidth]{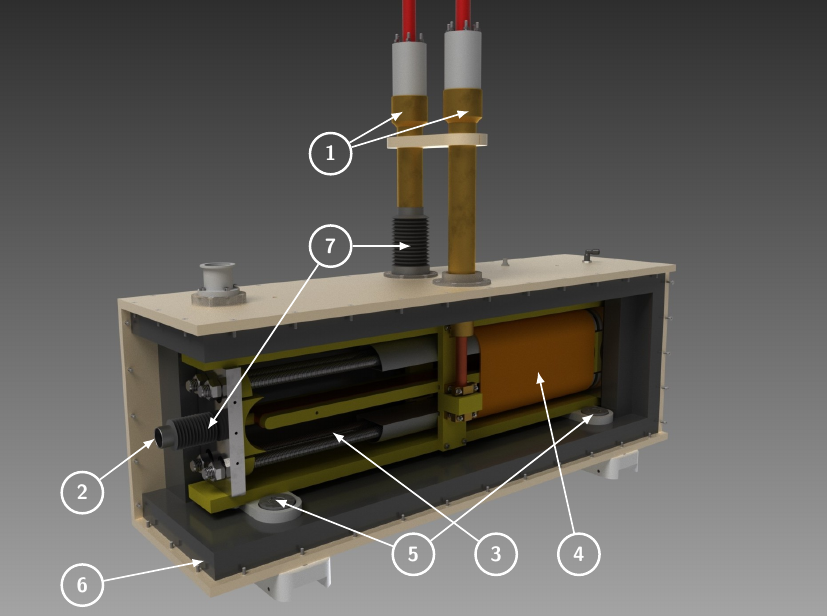}
\caption{Sketch of two cuts of the coil mechanical assembly of the Foil Coil used in the run 2025. 1: Two parallel current supplies. 2: Optical access provided by two holes in each turn of the insulated copper, to limit the coupling with the box a bellows is placed on each side between the aperture of the box and the winding. 3: Winding of copper foil insulated with two layers of Kapton tape. 4: Mechanical reinforcement with Zylon fibres. 5: Suspension springs. 6: Acoustic foam. 7: Insulation bellows composed of Neoprene.}\label{fig15}
\end{figure}
\begin{figure}[h!]
\centering
\includegraphics[width=\textwidth]{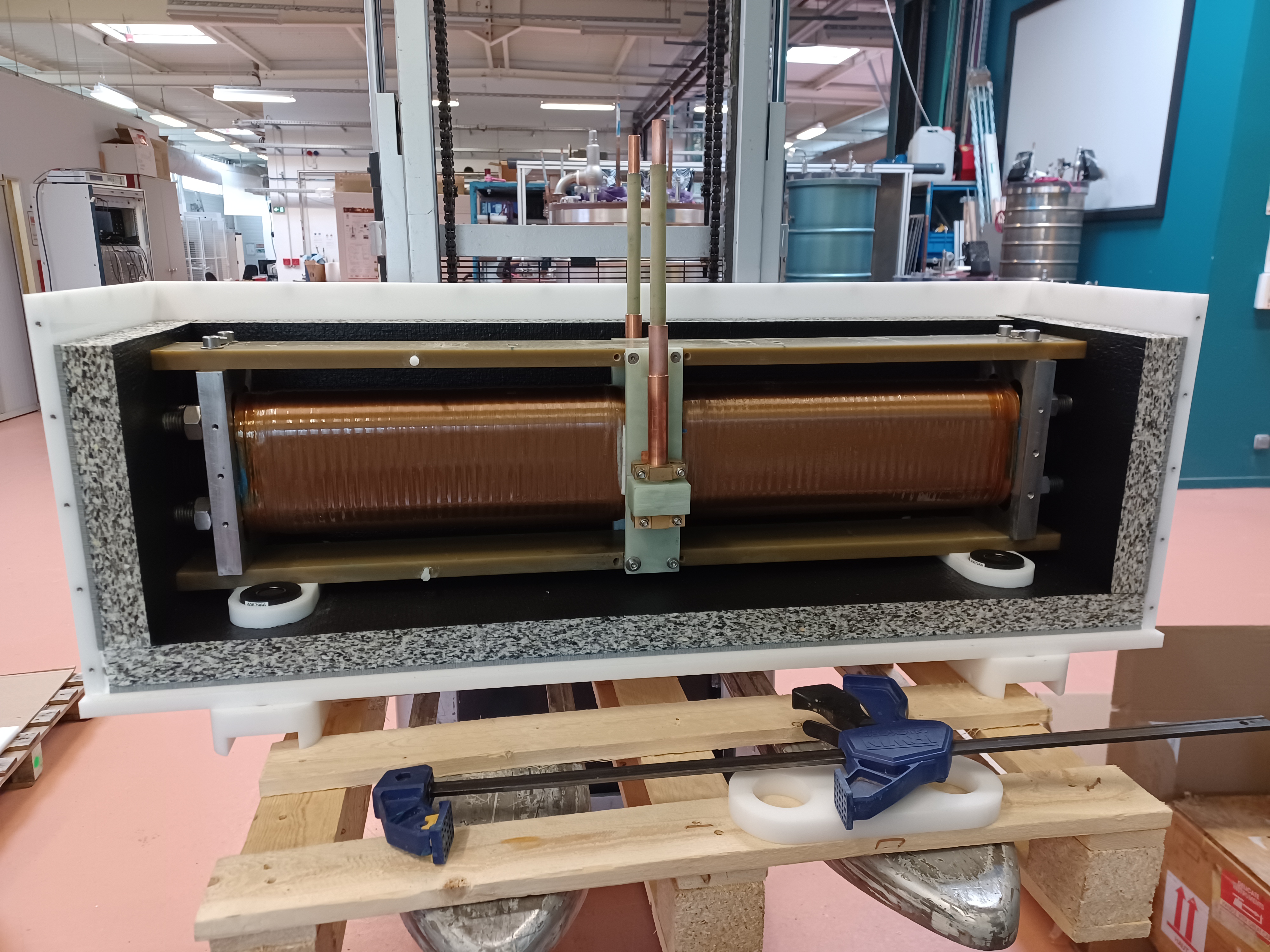}
\caption{Photo of the Foil Coil used in the run 2025 in its mechanical assembly.}\label{fig16}
\end{figure}
\begin{figure}[h!]
     \centering
     \begin{subfigure}[b]{\linewidth}
         \centering
         \includegraphics[width=\linewidth]{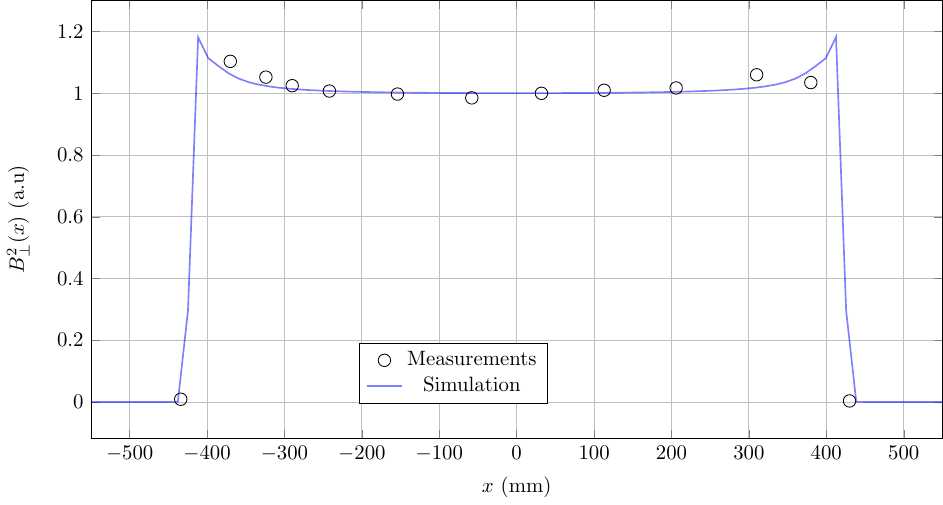}
         \caption{}
         \label{fig17a}
     \end{subfigure}
     \\
     \begin{subfigure}[b]{\linewidth}
         \centering
         \includegraphics[width=\linewidth]{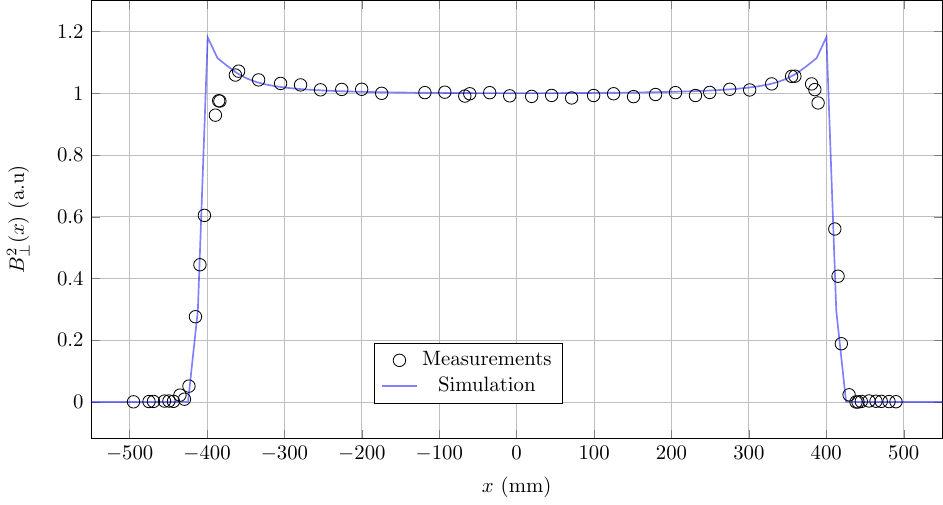}
         \caption{}
         \label{fig17b}
     \end{subfigure}
        \caption{Normalised profile of the magnetic field along the coil optical axis produced by the 2021 Foil Coil Fig.~\ref{fig17a} and 2025 Foil Coil Fig.~\ref{fig17b}. The blue solid line on each graph represents the values obtained with a finite element COMSOL simulation of the coil.}
        \label{fig17}
\end{figure}

As reported in ref.~\cite{Beard2021}, during a magnetic pulse the coil is also a source of acoustic noise. Thus, for run 2025 we have made some major modifications in the design of the coil with respect to run 2021. We doubled the current input (indicated as 1 in Fig. \ref{fig15}) to reduce the current in each of them and therefore reduce the force between them and the coil at the contact point. We have also introduced in the mechanical assembly a system of suspension and sound absorbers to reduce the impact of the sound generated during the pulse into the optical apparatus (see Fig.~\ref{fig15}~and~\ref{fig16}). 
\paragraph{Measurement of $B^2L_B$}
According to eq.~(\ref{PsiG}) to obtain a value of $K_{VMB}$ from a measurement of $\Psi(t)$ one needs to know the experimental parameter $B^2(t)L_B$. Thus, we performed during run 2021 and 2025 magnetic field amplitude measurements during a pulse by using a pickup coil inserted in the optical aperture of each Foil Coil. We show in Fig.~\ref{fig17} the $B^2(x)$ measurements as a function of the pickup coil position $x$. In Fig.~\ref{fig17} we also superimpose simulations by finite element. The obtained results, by normalising $B^2$ to 1~T$^2$, are $L_B \approx 0.86$~m in 2021 and $L_B \approx 0.83$~m in 2025.

\subsection{Calibration of the apparatus}
During run 2021, to fix the sign of $\Gamma$, coming from cavity mirror intrinsic birefringence, and calibrate the apparatus we have taken data with a pressure of a few tens mbar of helium gas.

During run 2025, thanks to the presence of the $\eta$ off-cavity wave-plate we have been able to calibrate the apparatus before installing the FP cavity, which is a technical novelty, measuring the N$_2$ gas Cotton-Mouton effect. 

\begin{table}[h!]
    \centering
\begin{tabular}{c c c}
\hline
\hline
     Parameter & Typical value & Relative uncertainty (\%) \\
     \hline
     $B^2L_B$ & 15.3 T$^2\cdot$m & 2 \\
     $\lambda$ & 1064 nm & $<0.05$\\
     $\sin{\left(2\theta\right)}$ & 1.000 & $<0.1$\\
     $P_e$ & 10$^{-10}$ W & 6 \\
     $P_t$ & 4$\cdot 10^{-3}$ W & 6 \\
     \hline
     & Total & 8.6 \% \\
     \hline
     \hline
\end{tabular}
\caption{Parameters used to perform $K_{CM}(P)$ measurements for 2025 apparatus calibration, a typical value and the relative uncertainties are shown for each parameters.}
    \label{tab1}
\end{table}

\begin{figure}[h!]
\centering
\includegraphics[width=\textwidth]{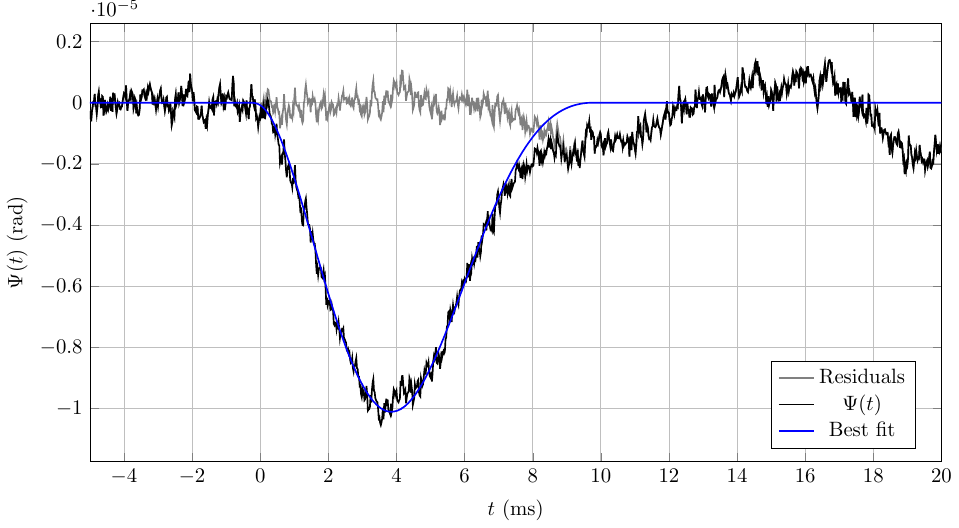}
\caption{Example of an ellipticity signal $\Psi(t)$ given by a quadruplet in $\text{N}_{2(\text{g})}$ for
a shot at $\approx 4.3 $~T. The best fit residuals are shown in light gray. The fit gives a
Cotton-Mouton constant $K_{CM} = (-2.18 \pm 0.19) \cdot 10^{-13}$~T$^{-2}$, the uncertainty takes into account the fitting and error budget uncertainties (see table \ref{tab1}). $\text{N}_{2(\text{g})}$ pressure $P = (0.92 \pm 0.02)$~atm.}\label{fig19}
\end{figure}

As shown by eq.~(\ref{PsiWOcav}), we have a fitting ellipticity $\Psi(t)$ function of time thanks to the modulated temporal profile of $B(t)$ and $\Psi(t)$ is proportional to $L_BB^2(t)$. As explained in \ref{subsec37} we need to exploit the symmetry properties with respect to the sign of $B$ and $\eta$ with 4 magnetic pulses in 4 different configurations to get the signal $\Psi^{(-+)}(t)$. We will name a series of four pulses a quadruplet. We show a $\Psi^{(-+)}(t)$ quadruplet with a best fit function superimposed between 0 and 8~ms in Fig.~\ref{fig19}. 

\begin{figure}[h!]
\centering
\includegraphics[width=\textwidth]{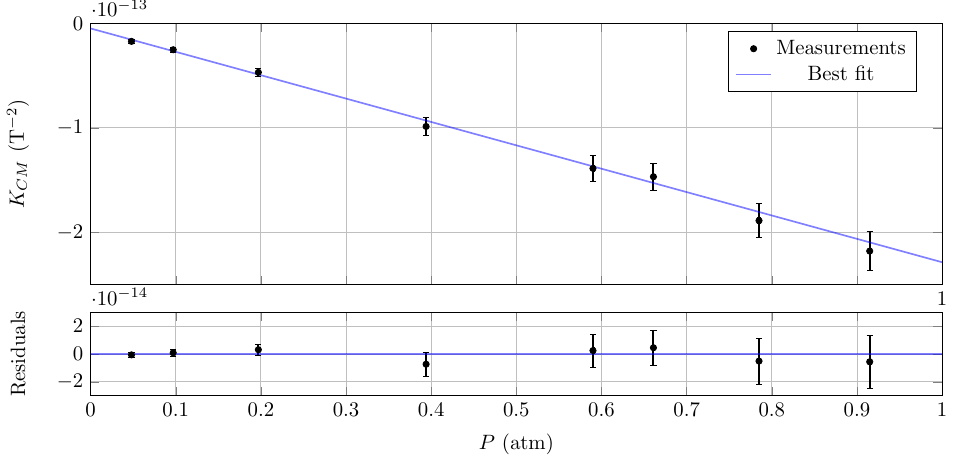}
\caption{Measurements of $K_{CM}$ in $\text{N}_{2(\text{g})}$ as a function of pressure. The best fit superimposed gives $\Delta n_u = (-2.24 \pm 0.10) \cdot 10^{-13}$~T$^{-2}\cdot$~atm$^{-1}$.}\label{fig20}
\end{figure}

\begin{figure}[h!]
\centering
\includegraphics[width=\textwidth]{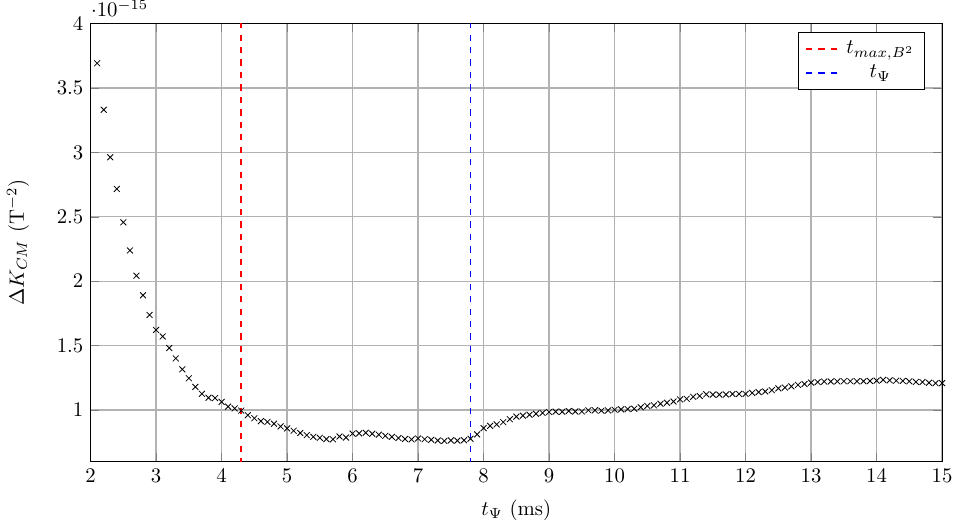}
\caption{Uncertainty $u\left(K_{CM}\right)$ as a function of the best fit analysis time $t_\Psi$. Red dashed line corresponds to the time when the function $B^2(t)$ is maximum. Blue dashed line corresponds to the time $t_\Psi$ chosen for the analysis.}
\label{fig21}
\end{figure}

As shown in Fig.~\ref{fig20}, we have measured different Cotton-Mouton constants $K_{CM}(P)$ in N$_{2(\text{g})}$ at different pressure $P$ in the range 0.05 atm to 1~atm. Assuming, as usual, that in this range of pressure the N$_{2(\text{g})}$ is an ideal gas, we can fit by $K_{CM}(P) = \Delta n_u P + b$ where $\Delta n_u$ is the Cotton-Mouton constant for $B^2(t) = 1$ T$^2$ and $P = 1$~atm. Our results as a function of gas pressure are shown in Fig.~\ref{fig20}. All obtained by the best fit performed until the time $t_\Psi$ corresponding to the moment when mechanical perturbations induced by the pulse itself affect the data. For example, in Fig.~\ref{fig19} one sees that $t_\Psi$ is about 7.5~ms. To evaluate $t_\Psi$ more precisely we plot the $K_{CM}(t_\Psi)$ uncertainty of the best fit itself as a function of $t_\Psi$ looking for the moment where the $u\left(K_{CM}(t_\Psi)\right)$ starts to increase as shown in Fig.~\ref{fig21} for the 0.65 atm signal.

The best fit gives as final result $\Delta n_u = (-2.24 \pm 0.10)\cdot 10^{-13}$ T$^{-2}\cdot$~atm$^{-1}$ with 68\% of confidence level.

As shown in table \ref{tab2}, our measurement of $\Delta n_u$ of N$_{2(\text{g})}$ agrees with other measurements published at $\lambda = 1064$ nm using an optical cavity to increase the effect and also with theoretical calculations of quantum chemistry as reported in \cite{Rizzo1997}.

\begin{table}[h!]
    \centering
\begin{tabular}{c c}
\hline
\hline
     References & $\Delta n_u \cdot 10^{13}$, $P = 1$ atm, $B = 1$ T \\
     \hline
     Bregant et al. \cite{bregant2004measurement} & (-2.17 \(\pm\) 0.21) \(\ \text{T}^{-2}.\text{atm}^{-1}\) \\
     H.-H. Mei et al. \cite{mei2009measurement} & (-2.02 \(\pm\) 0.18) \(\ \text{T}^{-2}.\text{atm}^{-1}\)\\
     Berceau et al. \cite{Berceau2012} & (-2.00 \(\pm\) 0.10) \(\ \text{T}^{-2}.\text{atm}^{-1}\)\\
     \hline
     This work & (-2.24 \(\pm\) 0.10) \(\ \text{T}^{-2}.\text{atm}^{-1}\)\\
     \hline
     \hline
\end{tabular}
\caption{Comparison of our measurement with others published at $\lambda = 1064$~nm and using an optical cavity.}
    \label{tab2}
\end{table}

\subsection{$K_{VMB}$ measurement results}
Since 2021, we have performed about 1200 pulses to establish our apparatus sensitivity in $K_{VMB}$. Pulses correspond to different maximum fields and different rising time to test the existence of any effect depending on the temporal derivative of $B(t)$.

A non negligible number of these pulses were affected by systematic noises and we had to define some criteria to take them out from the final analysis. The same has been done for the 2014 data reported in \cite{Cadene2014}, where 50\% of the total number of pulses were analysed notwithstanding a much more relaxed rejection criteria. 

After applying our rejection criteria to select the sets to be analysed, we have to choose an analysis time. Similarly to our calibration measurement, we first determine the time when the standard deviation of $\Psi^{(-+)}(t)$ begin to increase compared to the one before the beginning of the magnetic pulse. After that, we choose around this time, the $t_\Psi$ that gives us the $\chi^2_\nu$ of the best fit closest to 1.

\begin{figure}[h!]
    \centering
    \includegraphics[width=\textwidth]{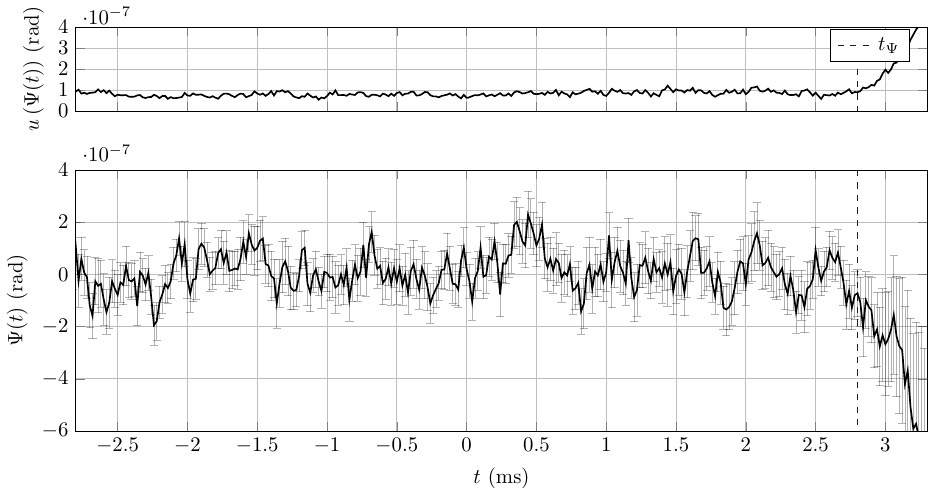}
    \caption{Below, the ellipticity signal $\Psi(t)$ of the 3.9S4 set from run 2025, above his statistical error-bar $u\left(\Psi(t)\right)$ with confidence level of 68\%. The chosen time analysis of this set is 2.8~ms and shown on each graph by the vertical dashed blue line.}
    \label{fig22}
\end{figure}

Let's note that each set is composed of $\mathrm{N}_\text{S}$ magnetic pulses in which the ellipticity signal is measured over a 50 ms length time at a sampling frequency of 50 kHz and finally computed according to eq.~(\ref{eq:TruePsi}). In Fig.~\ref{fig22}, we show the ellipticity $\Psi(t)$ of a set.

In the following, we present our $K_{VMB}$ measurements results of the runs 2021 and 2025 for each sets chosen, and finally we give a new limit on the $K_{VMB}$ from all selected sets. Let's first of all detail the rejection criteria used to analyse runs 2021 and 2025 data.

\subsubsection{Rejection criteria}
Our $K_{VMB}$ measurement has to be both precise and accurate. As far as precision is concerned, it can be increased by averaging as many sets as possible, but, as we discussed in \ref{Systeff}, this can affect the measurement accuracy if data containing spurious signals are taken into account. To take care therefore of our accuracy, we defined some criteria in order to eliminate spurious signal from the final average.

Even if there were some arbitrariness in the choosing of the sets to be eliminated, we consider that is more important to lose in precision than in accuracy.

\paragraph{Optical $K_{VMB}$ criterion}
Let's define $\sigma_\text{o}$, for a given set, the average value of the standard deviation of the ellipticity signal of each pulses on the time interval $-t_\Psi < t < 0$. We will call it in the following the optical noise. Thus, the best value $K_{VMB,o}$ of a given set can be written as
\begin{equation}
    K_{VMB,o} = \dfrac{\lambda}{2\sin{\left(2\theta\right)}L_B\mathbf{B}^2_f F\sqrt{\mathrm{N}_\text{S}}\sqrt{m}}\sigma_\text{o},
\end{equation}
where $m$ is the number of data points within the time $t_\Psi$ analysis duration, $\mathrm{N}_\text{S}$ the number of pulses composing the set. The rejection criteria applied to a set corresponding to the measurement $K_{VMB} \pm u\left(K_{VMB}\right)$ obtained by the best fit analysis was that $K_{VMB,o} <  u\left(K_{VMB}\right)$. The corresponding value $K_{VMB,o}$ for each sets is reported for the two runs on the table \ref{tab4} and \ref{tab6}.

\begin{figure}[h!]
    \centering
    \includegraphics[width=\textwidth]{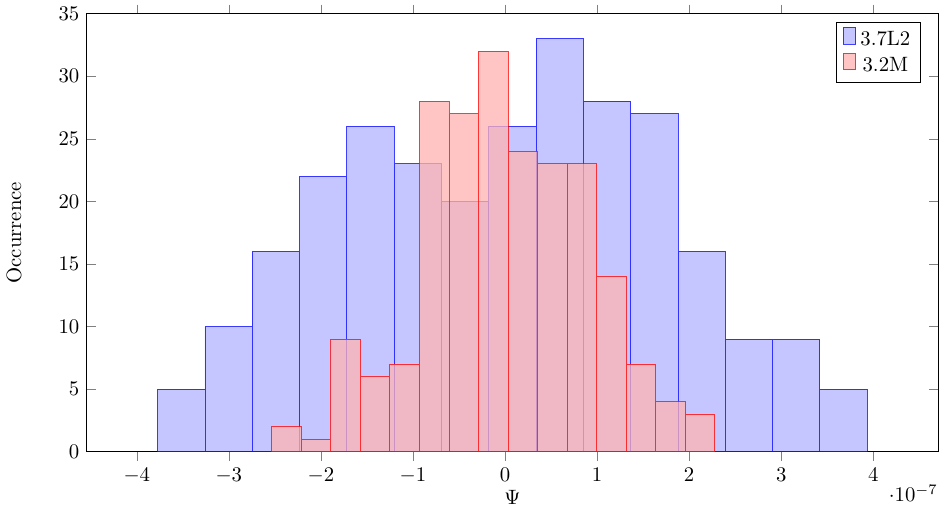}
    \caption{Histograms of the ellipticity signal $\Psi(t)$ before the beginning of the magnetic pulse ($-t_\Psi<t<0$). In red, we show a Gaussian distribution that we kept for analysis, and in blue a non-Gaussian distribution that we are rejecting.}
    \label{fig23}
\end{figure}

\paragraph{Gaussian noise criterion}
We also checked the histogram of the ellipticity signal for each set for $-t_\Psi<t<0$ and we reject the sets corresponding to a histogram that could be better fitted with a double Gaussian distribution instead of a single one, as shown in the example of the Fig.~\ref{fig23}. This double Gaussian distribution indicates that the value of $\Gamma$ cannot be properly fixed for each pulse, since the signal seems to having changed abruptly its average value just before the pulse.

This criterion was the only one applied for the analysis of our 2014 data \cite{Cadene2014} to each pulse of the only set.

\begin{figure}[h!]
\centering
\includegraphics[width=\textwidth]{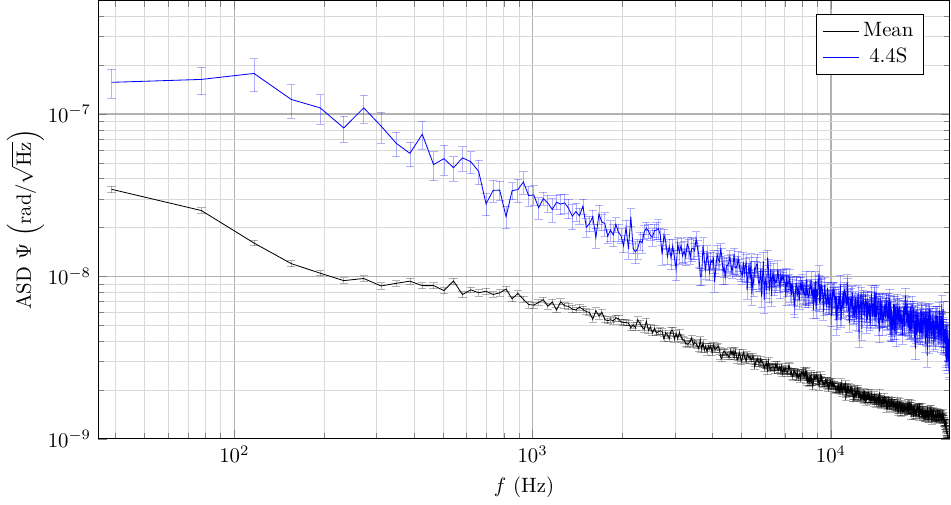}
\caption{In black we show the mean ASD of $\Psi(t)$ before the beginning of the magnetic pulse of the kept sets in run 2025 already shown in Fig.~\ref{fig18}. In blue we show the same quantity for a rejected set (4.4S).}
\label{fig24}
\end{figure}

\paragraph{ASD criterion}
The final criterion was based on the ASD of $\Psi(t)$ for $t<0$. Any set showing an ASD at more than 3 standard deviations from the average one obtained using the sets with the lowest ASD has been rejected. An example is shown in Fig.~\ref{fig24}.

\subsubsection{$K_{VMB}$ measurements, run 2021}
\begin{figure}[h!]
\centering
\includegraphics[width=1\textwidth]{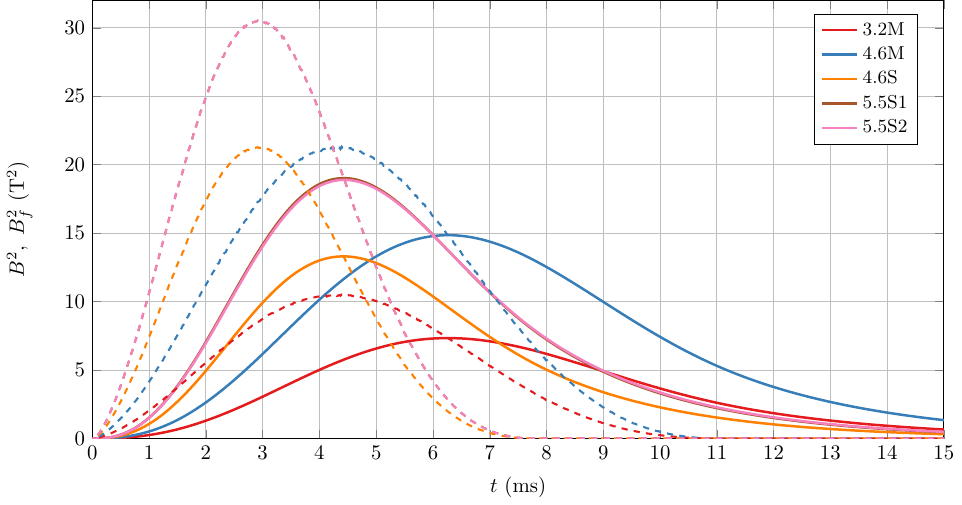}
\caption{All sets squared magnetic fields $B^2(t)$ in dashed lines and in solid lines the filtered ones $B^2_f(t)$ associated to them for the run 2021.}\label{fig25}
\end{figure}

We have performed 578 pulses distributed over 14 sets with three different pulse lengths and different pulse strengths, see Fig.~\ref{fig25}. The long, medium and short magnetic pulses have 5.8, 4.4 and 2.9~ms of rising time respectively. After applying our rejection criteria, we kept 141 pulses distributed over 5 sets.

We present in the table~\ref{tab3} the different sets and their main parameters obtained by the same method described in ref.~\cite{Beard2021}. The data sets are named according to their maximum $B(t)$ field and whether they are Medium (M) or Short (S) pulses. Pulses with the same field but on different days have been separated in different sets because we have to align the cavity each day to reach a static birefringence of the order of $\sigma^2$ and thus we are not in the same experimental conditions.

The best fit parameters are in table \ref{tab4}, the uncertainty $u\left(K_{VMB}\right)$ is given with 68\% level of confidence.

\begin{table}
\centering
\begin{tabular}{c c c c c c c }
\hline
\hline
Set & $\mathrm{N}_\text{S}$ & $B_f^2\ (\text{T}^2)$ & \(F\ \cdot 10^{-3}\) & \(\nu_{bw}\) (Hz) & \(\sigma^2\ \cdot 10^9\) & \(\left|\Gamma\right|\ (\cdot 10^5\ \text{rad})\) \\
\hline
3.2M & 51 & 7.3 & \(535 \pm 2\) & 55 & 1 & 5 \\
4.6M & 34 & 14.9 & \(537 \pm 2\) & 55 & 1 & 4 \\
4.6S & 15 & 13.3 & \(466 \pm 2\) & 63 & 16 & 12 \\
5.5S1 & 12 & 19.0 & \(469 \pm 2\) & 63 & 16 & 15 \\
5.5S2 & 29 & 18.9 & \(476 \pm 2\) & 62 & 4 & 7 \\
\hline
\hline
\end{tabular}
\caption{Experimental parameters of each set of 2021 run used for the analysis. The set marked with M or S corresponds to medium and short magnetic pulse length, respectively. $B_f^2$ correspond to the maximum of $B_f^2(t)$. $F$ is the cavity finesse of the corresponding set and $\nu_{bw}$ the cut-off frequency associated. $\sigma^2$ the extinction ratio and $\left|\Gamma\right|$ the static birefringence of the apparatus.}
\label{tab3}
\end{table}

\begin{table}
\centering
\begin{tabular}{c c c c c c c}
\hline
\hline
Set & $t_\Psi$ (ms) & \(\left(K_{VMB}\pm u\left(K_{VMB}\right)\right)\cdot 10^{20}\ \text{T}^{-2}\) & \(\chi^2_\nu\) & \(\sigma_\text{o}\) ($\cdot 10^6$rad) & \(K_{VMB,\text{o}}\cdot 10^{20}\ \text{T}^{-2}\)\\
\hline
3.2M & 4.2 & \(0.4 \pm 0.5\) & 0.90 & 1.3 & 0.05 \\
4.6M & 1.4 & \(-0.5 \pm 4.2\) & 1.11 & 1.8 & 3.44 \\
4.6S & 3.6 & \(-0.3 \pm 0.7\) & 1.08 & 1.8 & 0.03 \\
5.5S1 & 2.35 & \(0.8 \pm 0.7\) & 0.29 & 1.9 & 0.08 \\
5.5S2 & 3.3 & \(-0.9 \pm 0.4\) & 1.00 & 1.4 & 0.01 \\
\hline
\hline
\end{tabular}
\caption{Fitting parameters of the run 2021 sets, $\left(K_{VMB}\pm u\left(K_{VMB}\right)\right)$ is the VMB constant measurement with 68\% of level of confidence with the $\chi^2_\nu$ associated.}
\label{tab4}
\end{table}

\subsubsection{$K_{VMB}$ measurements, run 2025}
\begin{figure}[h!]
\centering
\includegraphics[width=1\textwidth]{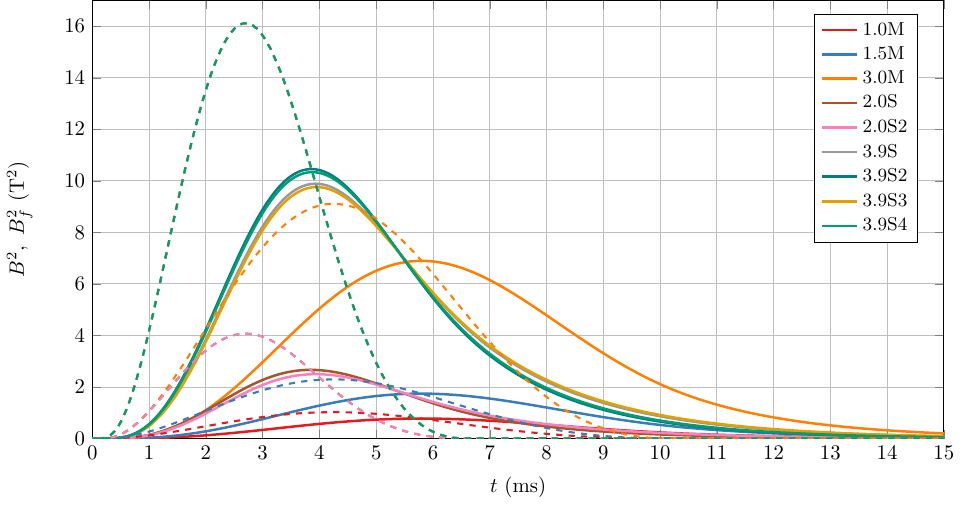}
\caption{All sets squared magnetic fields $B^2(t)$ in dashed lines and in solid lines the filtered ones $B^2_f(t)$ associated to them for the run 2025.}\label{fig26}
\end{figure}

As far as the run 2025 is concerned, we have performed 612 pulses distributed over 13 sets with two different pulse lengths and different pulse strengths, see Fig.~\ref{fig26}. The medium and short magnetic pulses have respectively 4.3 and 2.6~ms of rising time. After applying our rejection criteria, we kept 309 pulses distributed over 9 sets which corresponds to about 50\% of the total number of the acquired signals. It is an improvement with respect to the 24\% of the run 2021. As far as we understand, this is due to the use of the $\eta$ off-cavity wave-plate which allows us to align the cavity much more rarely between pulses during a set. This indicates that the cavity is in a much more stable situation than in 2021 where we had to turn and tilt mirrors to change the sign of $\Gamma$ and decrease $\sigma^2$.

We present in the table~\ref{tab5} the different sets and their experimental parameters. The data sets are named according to their maximum $B(t)$ field and whether they are Medium (M) or Short (S) pulses. For this run, by adding the off-cavity wave-plate $\eta$ the static birefringence is here $\left|\eta + \Gamma\right|$.

The best fit parameters are in table \ref{tab6}.

\begin{table}
\centering
\begin{tabular}{c c c c c c c }
\hline
\hline
Set & $\mathrm{N}_\text{S}$ & $B_f^2\ (\text{T}^2)$ & \(F\ \cdot 10^{-3}\) & \(\nu_{bw}\) (Hz) & \(\sigma^2\ \cdot 10^9\) & \(\left|\Gamma+\eta\right|\ (\cdot 10^5\ \text{rad})\) \\
\hline
1.0M & 20 & 0.8 & \(388 \pm 1\) & 76 & 10 & 16\\
1.5M & 20 & 1.7 & \(386 \pm 1\) & 76 & 10 & 17 \\
3.0M & 16 & 6.9 & \(390 \pm 1\) & 75 & 10 & 9 \\
2.0S & 17 & 2.7 & \(341 \pm 1\) & 86 & 30 & 20 \\
2.0S2 & 100 & 2.5 & \(393 \pm 1\) & 75 & 10 & 13 \\
3.9S & 53 & 9.9 & \(393 \pm 1\) & 75 & 6 & 18 \\
3.9S2 & 19 & 10.5 & \(348 \pm 1\) & 84 & 30 & 16 \\
3.9S3 & 30 & 9.8 & \(404 \pm 1\) & 73 & 30 & 13 \\
3.9S4 & 34 & 10.3 & \(357 \pm 1\) & 82 & 7 & 16 \\
\hline
\hline
\end{tabular}
\caption{Experimental parameters of each set from 2025 run used for the analysis. The set marked with M or S correspond to medium and short magnetic length respectively. $\mathrm{N}_\text{S}$ is the number of shots, $B_f^2$ value correspond to its maximum. $F$ is the cavity finesse of the corresponding set and $\nu_{bw}$ the cut-off frequency associated. $\sigma^2$ the extinction ratio and $\left|\Gamma+\eta\right|$ the static birefringence of the apparatus.}
\label{tab5}
\end{table}

\begin{table}
\centering
\begin{tabular}{c c c c c c c}
\hline
\hline
Set & $t_\Psi$ (ms) & \(\left(K_{VMB}\pm u\left(K_{VMB}\right)\right)\cdot 10^{20}\ \text{T}^{-2}\) & \(\chi^2_\nu\) & \(\sigma_\text{o}\) ($\cdot 10^6$rad) & \(K_{VMB,\text{o}}\cdot 10^{20}\ \text{T}^{-2}\) \\
\hline
1.0M & 6.7 & \(0.2 \pm 3.5\) & 0.95 & 0.7 & 2.63 \\
1.5M & 3.5 & \(2.6 \pm 3.3\) & 1.41 & 0.6 & 1.91 \\
3.0M & 1.9 & \(3.3 \pm 8.4\) & 1.33 & 0.8 & 4.24 \\
2.0S & 3.0 & \(0.0 \pm 1.6\) & 0.90 & 0.6 & 0.52 \\
2.0S2 & 3.7 & \(-4.7 \pm 1.3\) & 1.28 & 1.3 & 0.29 \\
3.9S & 2.3 & \(1.0 \pm 0.5\) & 0.86 & 0.6 & 0.05 \\
3.9S2 & 3.4 & \(0.0 \pm 0.5\) & 1.82 & 0.6 & 0.02 \\
3.9S3 & 3.3 & \(0.2 \pm 0.6\) & 0.98 & 0.9 & 0.03 \\
3.9S4 & 2.8 & \(-0.2 \pm 0.4\) & 0.83 & 0.6 & 0.03 \\
\hline
\hline
\end{tabular}
\caption{Fitting parameters of the run 2025 sets, $\left(K_{VMB}\pm u\left(K_{VMB}\right)\right)$ is the VMB constant measurement with 68\% of level of confidence with the $\chi^2_\nu$ associated.}
\label{tab6}
\end{table}

\subsection{Final $\langle K_{VMB} \rangle$ result}
\begin{figure}[h!]
    \centering
    \includegraphics[width=\textwidth]{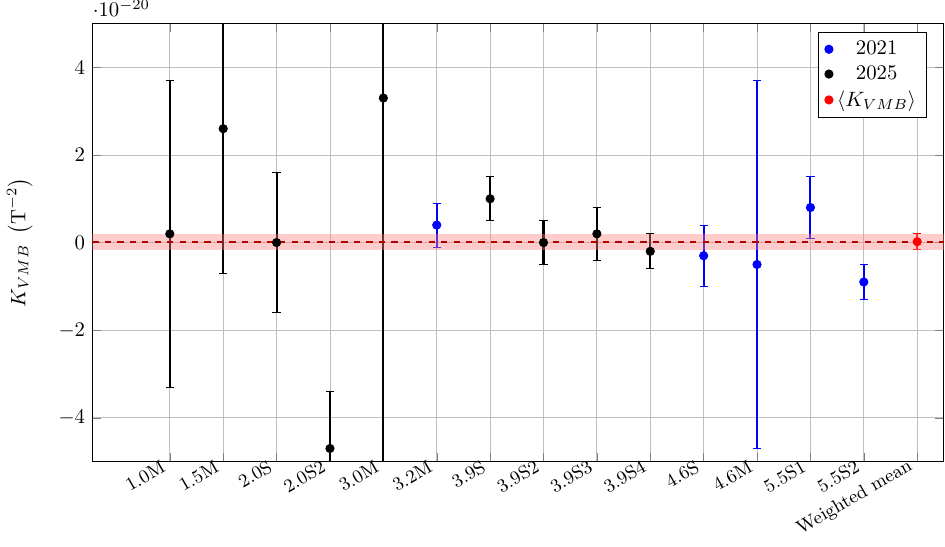}
    \caption{$K_{VMB}$ measurements of each set from both runs.}
    \label{fig27}
\end{figure}

All the $K_{VMB}$ measurements from the table~\ref{tab4} and \ref{tab6} are shown on the Fig.~\ref{fig27}.
In Fig.~\ref{fig27}, we also show the weighted mean of the measurements $\langle K_{VMB} \rangle \pm u\left(K_{VMB}\right)$ calculated by using the following formulas
\begin{align}
\langle K_{VMB} \rangle & = \dfrac{\sum_i^{\text{sets}} \frac{K_{VMB,i}}{\left(u\left(K_{VMB,i}\right)\right)^2} }{\sum_i^{\text{sets}} \frac{1}{\left(u\left(K_{VMB,i}\right)\right)^2}} = -0.6 \cdot 10^{-21} \ \text{T}^{-2} \\
u\left(K_{VMB}\right) & = \dfrac{1}{\sqrt{\sum_i^{\text{sets}}\frac{1}{\left(u\left(K_{VMB,i}\right)\right)^2}}} = 1.7 \cdot 10^{-21} \ \text{T}^{-2}\\
\chi^2_\text{red} & = \frac{1}{N-1}\sum_i^{\text{sets}} \frac{\left(K_{VMB,i}-\langle K_{VMB} \rangle\right)^2}{u\left(K_{VMB,i}\right)^2}= 1.9, 
\end{align}
where $N=14$ is the number of the sets used in the analysis.

Let's note that, if we do not take into account the set 2.0S2 which is clearly an outlier but that passes all the criteria, we obtain $\langle K_{VMB} \rangle = 0.2\cdot 10^{-21}$~T$^{-2}$, $u\left(K_{VMB}\right) = 1.8\cdot 10^{-21}$~T$^{-2}$ and $\chi^2_\text{red}= 1.0$.

\section{Conclusions and perspectives}\label{sec4}

\begin{figure}[h!]
    \centering
    \includegraphics[width=\textwidth]{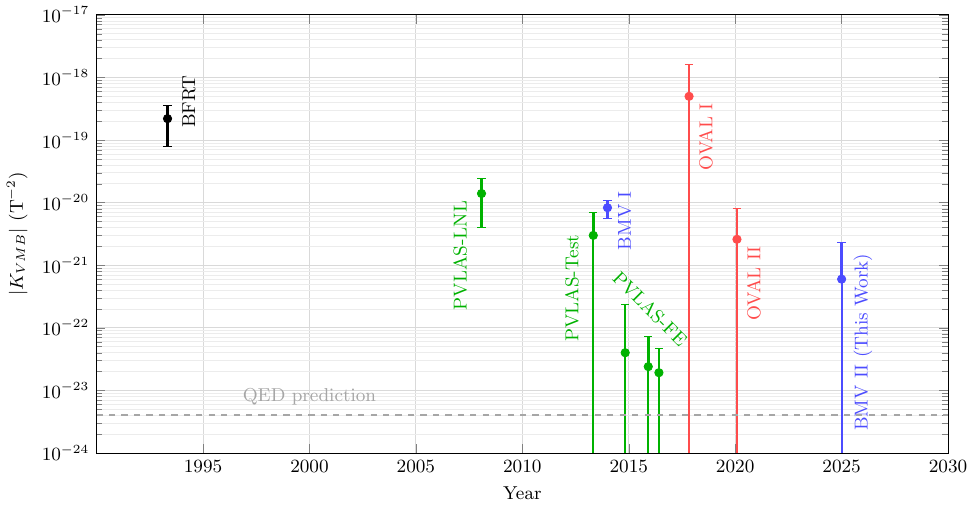}
    \caption{Measurements of the vacuum Cotton-Mouton constant $K_{VMB}$ across the years, errors bars are represented at 68\% level of confidence. Absolute values are derived from data originally published in the following references: BFRT\cite{Cameron1993}, PVLAS-LNL\cite{zavattini2008new}, PVLAS-Test\cite{della2013measurements}, BMV\cite{Cadene2014}, PVLAS-FE\cite{Ejlli2020}\cite{della2016pvlas}\cite{PhysRevD.90.092003}, OVAL\cite{fan2017oval}\cite{kamioka2020}.}
    \label{fig28}
\end{figure}

In this paper, we report on the data taken in recent years in the framework of the BMV project hosted in LNCMI in Toulouse, France. The novel final limit with a confidence level of 68\% on the vacuum magnetic birefringence constant that we have obtained is $$K_{VMB} = \left(-0.6 \pm 1.7\right)\cdot 10^{-21}\ \text{T}^{-2}.$$ This value is a clear improvement with respect to the 2014 value reported in \cite{Cadene2014} by our own group. The improvement concerns both accuracy and precision, see Fig.~\ref{fig28}. Let's also note 2025 results are also an improvement with respect to 2021 ones.

On the other hand, as also shown in Fig.~\ref{fig28}, using pulsed fields we have not yet been able to reach the best existing limit given by the PVLAS collaboration after more than 25 years of works \cite{Ejlli2020}. Let's note that this kind of experiment looks quite challenging and, as far as we know, our group is currently the only one remaining taking data to observe the vacuum magnetic birefringence for the first time.

Anyway, it is clear that we need further improvements to increase analysis time and therefore exploiting the entire pulse by reducing the perturbations induced by the pulse itself. This will naturally increase our sensitivity in $\Psi$ because we have a longer analysis time and in $\Delta n_{VMB}$ because we will be able to use the maximum field of each pulse and hopefully more than 10 T as it is possible with our Foil Coil. Last but not least, we have to also reduce the optical noise. This also will reduce the percentage of sets rejected increasing the number of pulses that will be taken into account with a corresponding increase of sensitivity. Increasing the field, for example, to 12.5~T will decrease the pulse duty cycle to about 1 pulse every 50 minutes \citep{Beard2021} which corresponds to a loss of a factor of $\sqrt{\frac{10}{60}/\frac{1}{50}} \approx 3$ on the noise for a fixed acquisition time and pulse duration but it also corresponds to a factor about 6 on the signal because of the increase of $B^2$.

 For all these reasons, we are currently designing a novel optical setup that will be operational as soon as possible. 

\section{Acknowledgements} 

We thank all the technical staff of the LNCMI, in particular Nicolas Bruyant, No{\'e} Estorges, Franck Giquel, Jean-Marc Lagarrigue. We also thank Julie Thomas-Matas a bachelor student, for her contribution to the $L_B$ measurement, and the former member of the BMV experiment M. T. Hartman, who greatly contributed to the design of the apparatus.
We kindly thank Pierre Lassegues for the time spent carefully reviewing this paper. 

\section{Conflict of interests}
The authors have nothing to disclose.

\section{Funding}
J. A. is grateful for the support of the Humboldt Foundation.

\section{Data availability statement}
The datasets generated during and/or analysed during the
current study are available from the corresponding author
on reasonable request.

\section{Author contribution statement}
All the authors were involved in the preparation of the
manuscript. All the authors have read and approved the
final manuscript.

\bibliography{BMVbibliography}
\end{document}